\documentclass[a4paper]{article} 
\usepackage{geometry}
			 
\usepackage{tgpagella}
\usepackage{fontawesome5}
\usepackage[usenames,dvipsnames,svgnames,table]{xcolor}
\definecolor{mybg}{HTML}{282A36}
\definecolor{mycl}{HTML}{44475A}
\definecolor{myfg}{HTML}{F8F8F2}
\definecolor{mycomment}{HTML}{6272A4}
\definecolor{mycyan}{HTML}{8BE9FD}
\definecolor{mygreen}{HTML}{50FA7B}
\definecolor{myorange}{HTML}{FFB86C}
\definecolor{mypink}{HTML}{FF79C6}
\definecolor{mypurple}{HTML}{BD93F9}
\definecolor{myred}{HTML}{FF5555}
\definecolor{myyellow}{HTML}{F1FA8C}
\definecolor{VividPurple}{HTML}{3E0097}
\definecolor{SlateGrey}{HTML}{2E2E2E}
\definecolor{LightGrey}{HTML}{666666}
\definecolor{DarkPastelRed}{HTML}{450808}
\definecolor{PastelRed}{HTML}{8F0D0D}
\definecolor{GoldenEarth}{HTML}{E7D192}
\definecolor{awesome-emerald}{HTML}{00A388}
\definecolor{awesome-emerald-dark}{HTML}{00806A} 
\definecolor{awesome-skyblue}{HTML}{0395DE}
\definecolor{awesome-skyblue-dark}{HTML}{0376B0}
\definecolor{awesome-red}{HTML}{DC3522}
\definecolor{awesome-red-dark}{HTML}{B02A1C}
\definecolor{awesome-pink}{HTML}{EF4089}
\definecolor{awesome-pink-dark}{HTML}{EC136D}
\definecolor{awesome-orange}{HTML}{FF6138}
\definecolor{awesome-orange-dark}{HTML}{FF3300}
\definecolor{awesome-nephritis}{HTML}{27AE60}
\definecolor{awesome-nephritis-dark}{HTML}{219150}
\definecolor{awesome-concrete}{HTML}{95A5A6}
\definecolor{awesome-concrete-dark}{HTML}{74898B}
\definecolor{awesome-darknight}{HTML}{131A28}
\definecolor{awesome-darknight-dark}{HTML}{101623}
\definecolor{awesome-snowwhite}{HTML}{F9FBFD}
\definecolor{awesome-snowwhite-dark}{HTML}{F3F6FB}
\definecolor{awesome-blue-dark}{HTML}{0000FF}
\definecolor{awesome-golden}{HTML}{E1AD21}
\definecolor{awesome-silver}{HTML}{AAA9AD}
\definecolor{darktext}{HTML}{414141}
\definecolor{darktext-dark}{HTML}{262626}
\definecolor{text}{HTML}{333333}
\definecolor{graytext}{HTML}{5D5D5D}
\definecolor{lighttext}{HTML}{999999}
\definecolor{VividRed}{HTML}{7e2635}
\definecolor{DarkRed}{HTML}{a5402d}
\definecolor{SlateGrey}{HTML}{2E2E2E}
\definecolor{LightGrey}{HTML}{666666}
\UseRawInputEncoding
\usepackage[english]{babel}

\usepackage{lipsum}
\usepackage{amsmath, nccmath} 

\usepackage{gensymb}
\usepackage{amssymb}
\usepackage{graphics}
\usepackage{graphicx}
\usepackage{epsfig}
\usepackage{fancyhdr}
\usepackage{wasysym}
\usepackage{latexsym}
\usepackage{amsfonts}

\usepackage{float}
\usepackage{color,colortbl}
\usepackage{cite}
\usepackage{notoccite} 
\usepackage{soul}
\usepackage{multirow}
\usepackage{etoolbox}
\usepackage{authblk}
\usepackage{blindtext}
\usepackage{dcolumn}
\usepackage{bm}
\usepackage{qcircuit}
\usepackage{braket}
\usepackage{physics}
\usepackage{braket}				   
\usepackage{varwidth}
\usepackage{empheq}
\usepackage[]{algorithm2e}
						   
\usepackage{algorithmic}
\usepackage{framed}
\usepackage{enumitem}
\usepackage[utf8x]{inputenc} 
\usepackage[english]{babel}
\usepackage{url}
\usepackage{hyperref}
\usepackage[font={footnotesize}]{caption}
\usepackage{enumitem}
\usepackage{multirow}
\usepackage{textcomp}

\usepackage[T1]{fontenc}
\usepackage{listings}
\usepackage{subcaption}
\usepackage{textcomp}
\usepackage{mathrsfs}
\usepackage{hyperref}
\usepackage{caption}
\usepackage{cleveref}
\usepackage{xr}
\graphicspath{{./figures}}
\usepackage{amsthm}
\usepackage{enumerate}
\usepackage{array}
\usepackage[parfill]{parskip}
\usepackage{amscd}
\usepackage[]{units}
\usepackage{multicol}
\usepackage{tcolorbox}
\usepackage{media9}
\usepackage{pifont}
\usepackage{booktabs}
\usepackage[acronym]{glossaries}
\usepackage{alphalph}
\usepackage{mathtools}
\usepackage{blindtext}
\usepackage{subfiles}
\usepackage{fullpage}
\usepackage{makeidx}
\usepackage{fancyhdr}
\usepackage[calcwidth]{titlesec}
\usepackage[font={footnotesize},labelfont={bf}]{caption}
\usepackage{setspace}%
\providecommand{\U}[1]{\protect\rule{.1in}{.1in}}
\usepackage[symbol]{footmisc}
\usepackage{bookmark}
\usepackage[compat=1.1.0]{tikz-feynman}
\usepackage{tikz}

\usetikzlibrary{spy,calc}
\usetikzlibrary{matrix,backgrounds}
\usetikzlibrary{shapes.geometric, arrows}
\usetikzlibrary{decorations.pathmorphing}
\tikzstyle{startstop} = [rectangle, rounded corners, minimum width=3cm, minimum height=1cm,text centered, draw=black, fill=red!30]
\tikzstyle{io} = [trapezium, trapezium left angle=70, trapezium right angle=110, minimum width=3cm, minimum height=1cm, text centered, draw=black, fill=blue!30]
\tikzstyle{process} = [rectangle, minimum width=3cm, minimum height=1cm, text centered, text width=3cm, draw=black, fill=orange!30]
\tikzstyle{decision} = [diamond, minimum width=3cm, minimum height=1cm, text centered, draw=black, fill=green!30]
\tikzstyle{arrow} = [thick,->,>=stealth]

\tikzstyle{decision} = [diamond, draw, fill=blue!20, 
text width=4.5em, text badly centered, node distance=3cm, inner sep=0pt]
\tikzstyle{block} = [rectangle, draw, fill=blue!20, 
text width=5em, text centered, rounded corners, minimum height=4em]
\tikzstyle{line} = [draw, -latex']
\tikzstyle{cloud} = [draw, ellipse,fill=red!20, node distance=3cm,
minimum height=2em]		

\usetikzlibrary{shadows, patterns}
\tikzset{button/.style={
preaction={fill=blue,path fading=circle with fuzzy edge 20 percent,
opacity=.7,transform canvas={xshift=1mm,yshift=-1mm}},
preaction={pattern=#1,
path fading=circle with fuzzy edge 20 percent},
preaction={top color=white,
bottom color=red!50,
shading angle=180,
path fading=circle with fuzzy edge 20 percent,
opacity=0.4},
preaction={path fading=fuzzy ring 15 percent,
top color=black!5,
bottom color=black!80,
shading angle=180},
inner sep=2ex
},
button/.default=horizontal lines light blue,
circle}	
\newif\ifblackandwhitecycle
\gdef\patternnumber{0}

\pgfkeys{/tikz/.cd,
zoombox paths/.style={
draw=orange,
very thick
},
black and white/.is choice,
black and white/.default=static,
black and white/static/.style={ 
draw=white,   
zoombox paths/.append style={
draw=white,
postaction={
draw=black,
loosely dashed
}
}
},
black and white/static/.code={
\gdef\patternnumber{1}
},
black and white/cycle/.code={
\blackandwhitecycletrue
\gdef\patternnumber{1}
},
black and white pattern/.is choice,
black and white pattern/0/.style={},
black and white pattern/1/.style={    
draw=white,
postaction={
draw=black,
dash pattern=on 2pt off 2pt
}
},
black and white pattern/2/.style={    
draw=white,
postaction={
draw=black,
dash pattern=on 4pt off 4pt
}
},
black and white pattern/3/.style={    
draw=white,
postaction={
draw=black,
dash pattern=on 4pt off 4pt on 1pt off 4pt
}
},
black and white pattern/4/.style={    
draw=white,
postaction={
draw=black,
dash pattern=on 4pt off 2pt on 2 pt off 2pt on 2 pt off 2pt
}
},
zoomboxarray inner gap/.initial=5pt,
zoomboxarray columns/.initial=2,
zoomboxarray rows/.initial=2,
subfigurename/.initial={},
figurename/.initial={zoombox},
zoomboxarray/.style={
execute at begin picture={
\begin{scope}[
spy using outlines={%
zoombox paths,
width=\imagewidth / \pgfkeysvalueof{/tikz/zoomboxarray columns} - (\pgfkeysvalueof{/tikz/zoomboxarray columns} - 1) / \pgfkeysvalueof{/tikz/zoomboxarray columns} * \pgfkeysvalueof{/tikz/zoomboxarray inner gap} -\pgflinewidth,
height=\imageheight / \pgfkeysvalueof{/tikz/zoomboxarray rows} - (\pgfkeysvalueof{/tikz/zoomboxarray rows} - 1) / \pgfkeysvalueof{/tikz/zoomboxarray rows} * \pgfkeysvalueof{/tikz/zoomboxarray inner gap}-\pgflinewidth,
magnification=3,
every spy on node/.style={
zoombox paths
},
every spy in node/.style={
zoombox paths
}
}
]
},
execute at end picture={
\end{scope}
\node at (image.north) [anchor=north,inner sep=0pt] {\subcaptionbox{\label{\pgfkeysvalueof{/tikz/figurename}-image}}{\phantomimage}};
\node at (zoomboxes container.north) [anchor=north,inner sep=0pt] {\subcaptionbox{\label{\pgfkeysvalueof{/tikz/figurename}-zoom}}{\phantomimage}};
\gdef\patternnumber{0}
},
spymargin/.initial=0.5em,
zoomboxes xshift/.initial=1,
zoomboxes right/.code=\pgfkeys{/tikz/zoomboxes xshift=1},
zoomboxes left/.code=\pgfkeys{/tikz/zoomboxes xshift=-1},
zoomboxes yshift/.initial=0,
zoomboxes above/.code={
\pgfkeys{/tikz/zoomboxes yshift=1},
\pgfkeys{/tikz/zoomboxes xshift=0}
},
zoomboxes below/.code={
\pgfkeys{/tikz/zoomboxes yshift=-1},
\pgfkeys{/tikz/zoomboxes xshift=0}
},
caption margin/.initial=4ex,
},
adjust caption spacing/.code={},
image container/.style={
inner sep=0pt,
at=(image.north),
anchor=north,
adjust caption spacing
},
zoomboxes container/.style={
inner sep=0pt,
at=(image.north),
anchor=north,
name=zoomboxes container,
xshift=\pgfkeysvalueof{/tikz/zoomboxes xshift}*(\imagewidth+\pgfkeysvalueof{/tikz/spymargin}),
yshift=\pgfkeysvalueof{/tikz/zoomboxes yshift}*(\imageheight+\pgfkeysvalueof{/tikz/spymargin}+\pgfkeysvalueof{/tikz/caption margin}),
adjust caption spacing
},
calculate dimensions/.code={
\pgfpointdiff{\pgfpointanchor{image}{south west} }{\pgfpointanchor{image}{north east} }
\pgfgetlastxy{\imagewidth}{\imageheight}
\global\let\imagewidth=\imagewidth
\global\let\imageheight=\imageheight
\gdef\columncount{1}
\gdef\rowcount{1}

},
image node/.style={
inner sep=0pt,
name=image,
anchor=south west,
append after command={
[calculate dimensions]
node [image container,subfigurename=\pgfkeysvalueof{/tikz/figurename}-image] {\phantomimage}
node [zoomboxes container,subfigurename=\pgfkeysvalueof{/tikz/figurename}-zoom] {\phantomimage}
}
},
color code/.style={
zoombox paths/.append style={draw=#1}
},
connect zoomboxes/.style={
spy connection path={\draw[draw=none,zoombox paths] (tikzspyonnode) -- (tikzspyinnode);}
},
help grid code/.code={
\begin{scope}[
x={(image.south east)},
y={(image.north west)},
font=\footnotesize,
help lines,
overlay
]
\foreach \x in {0,1,...,9} { 
\draw(\x/10,0) -- (\x/10,1);
\node [anchor=north] at (\x/10,0) {0.\x};
}
\foreach \y in {0,1,...,9} {
\draw(0,\y/10) -- (1,\y/10);                        \node [anchor=east] at (0,\y/10) {0.\y};
}
\end{scope}    
},
help grid/.style={
append after command={
[help grid code]
}
},
}

\newcommand\phantomimage{%
\phantom{%
\rule{\imagewidth}{\imageheight}%
}%
}
\newcommand\zoombox[2][]{
\begin{scope}[zoombox paths]
\pgfmathsetmacro\xpos{
(\columncount-1)*(\imagewidth / \pgfkeysvalueof{/tikz/zoomboxarray columns} + \pgfkeysvalueof{/tikz/zoomboxarray inner gap} / \pgfkeysvalueof{/tikz/zoomboxarray columns} ) + \pgflinewidth
}
\pgfmathsetmacro\ypos{
(\rowcount-1)*( \imageheight / \pgfkeysvalueof{/tikz/zoomboxarray rows} + \pgfkeysvalueof{/tikz/zoomboxarray inner gap} / \pgfkeysvalueof{/tikz/zoomboxarray rows} ) + 0.5*\pgflinewidth
}
\edef\dospy{\noexpand\spy [
#1,
zoombox paths/.append style={
black and white pattern=\patternnumber
},
every spy on node/.append style={#1},
x=\imagewidth,
y=\imageheight
] on (#2) in node [anchor=north west] at ($(zoomboxes container.north west)+(\xpos pt,-\ypos pt)$);}
\dospy
\pgfmathtruncatemacro\pgfmathresult{ifthenelse(\columncount==\pgfkeysvalueof{/tikz/zoomboxarray columns},\rowcount+1,\rowcount)}
\global\let\rowcount=\pgfmathresult
\pgfmathtruncatemacro\pgfmathresult{ifthenelse(\columncount==\pgfkeysvalueof{/tikz/zoomboxarray columns},1,\columncount+1)}
\global\let\columncount=\pgfmathresult
\ifblackandwhitecycle
\pgfmathtruncatemacro{\newpatternnumber}{\patternnumber+1}
\global\edef\patternnumber{\newpatternnumber}
\fi
\end{scope}
}				
\definecolor{anti-flashwhite}{rgb}{0.95, 0.95, 0.96}
\definecolor{codegreen}{rgb}{0,0.6,0}
\definecolor{codepurple}{rgb}{0.58,0,0.82}
\lstdefinelanguage{NeMO}{
keywords={},
ndkeywords={solver},
keywordstyle=\color{blue},
ndkeywordstyle=\color{codepurple},
commentstyle=\color{codegreen},
stringstyle=\color{cyan},
sensitive=true
}
\hypersetup{colorlinks=true,
linktocpage=true,
colorlinks=true,
linkcolor=awesome-blue-dark,
filecolor=awesome-red,
citecolor=awesome-pink-dark,      
urlcolor=awesome-orange-dark,
pdftitle={Lagrangian~Density},
bookmarks=true,
pdfauthor={Bhupesh~Bishnoi},
bookmarks=true,			   
bookmarksopen=true,
pdfpagemode=UseOutlines,
pdfstartview={FitH},	
unicode   =true,
pdftoolbar=true,
pdfmenubar=true,
}
\makeatletter
\def\@fnsymbol#1{\ensuremath{\ifcase#1\or \pmb\ddagger \else\@ctrerr\fi}}
\makeatother

\title {\Huge {Lagrangian Density Space-Time Deep Neural Network Topology}}
\author{\bf{Bhupesh~Bishnoi\thanks{\href{mailto:bishnoi@computer.org}{\textcolor{PastelRed}{\texttt{\bf{bishnoi[At]computer[Dot]org}}}}~~~\href{mailto:bishnoi@ieee.org}{\textcolor{PastelRed}{\texttt{\bf{bishnoi[At]ieee[Dot]org}}}} }}}
\date{}
\begin{document}

\maketitle

\begin{abstract} 
As a network-based functional approximator, we have proposed a ``Lagrangian Density Space-Time Deep Neural Networks'' (LDDNN) topology. It is qualified for unsupervised training and learning to predict the dynamics of underlying physical science governed phenomena. The prototypical network respects the fundamental conservation laws of nature through the succinctly described Lagrangian and Hamiltonian density of the system by a given data-set of generalized nonlinear partial differential equations. The objective is to parameterize the Lagrangian density over a neural network and directly learn from it through data instead of hand-crafting an exact time-dependent ``Action solution'' of Lagrangian density for the physical system. With this novel approach, can understand and open up the information inference aspect of the ``Black-box deep machine learning representation''  for the physical dynamics of nature by constructing custom-tailored network interconnect topologies, activation, and loss/cost functions based on the underlying physical differential operators. This article will discuss statistical physics interpretation of neural networks in the Lagrangian and Hamiltonian domains.	
\end{abstract}	
\section*{\texorpdfstring{Introduction}{Introduction}}
Our progress in understanding the nature from the pre-eighteenth-century experimentation approach to mid-twentieth-century physical theory-based modeling, model-based scientific computation, and first-principles simulation into today changing towards the fusion of data-driven deep learning and physical modeling simulation and statistical optimization. \cite{kinzel_statistical_1999,10.1145/3448250}
In the past decade, the neural network remarkably grew with the success of MNIST \cite{726791} and ImageNet projects, \cite{ImageNet_VSS09}, especially in the image processing and recognition field. The result is evident in image processing, robotics, and automation by observing the final resultant outcome. However, in the natural science physics, chemistry, material science domain, to evaluate the performance of the ``Black box deep neural network'',  we have to interpret both the algorithm and the internal state and the description of hidden layers node such as neural network potential. This article will investigate and emphasize the physical aspect of the deep neural network and correlate it with the informational domain of statistical inference.

Von Neuman first proposed that neural networks in the human biological cortex function as random digital machines and a stochastic treatment of their dynamics. \cite{von2012computer} From the ergodic nature of the system, we use satistical inference to predict the dynamical solution of the system, which is beyond the physical theory of the equation of motion that governs the dynamics. 
\cite{jaynes_information_1957,jaynes_information_1957-1,little1974existence,little1978analytic,hopfield1982neural,ingber1982statistical,PhysRevE.55.4578}
The steady mathematical correspondence between dynamical equation and stochastic evolution of the system derives through the fluctuation-dissipation theorem in physics.  However, the fluctuation-dissipation theorem is much deeper, more generalized, and associated with the subjective probability of the information of nature that we ever or should ask, and able to measure from nature and physical reality of observation associated with the measurables in an experiment. \cite{onsager_reciprocal_1931, onsager_reciprocal_1931-1,PhysRev.91.1505,PhysRev.91.1512} 
High-level deep learning (DL), machine learning (ML) application programming interface (API) libraries in deep neural networks implement predictive statistical mechanics. By the information reasoning process, probability theory mathematically tries to deduce the best prediction about the observable phenomena given the partial information about the state. It understands as physical analogous to the quantum state measurement technique. We try to deduce the system state through another set of measurements or observations as a pure quantum state does not contain enough information to predict all possible future outcomes. By this predictive inference, statistical physics also does not claim absolute certainty of deduction; however, it ensures that predictions are objective and forbid the extraneous assumption on the available dataset at the learning, training, validating, and testing stage. At this point, it is worth mentioning that those predictive outcomes may or may not succeed as a natural dynamical evolution of the quantum state. However, after going through the process in both fail-pass scenarios, We will know far more about the quantum state than we know in our present state. Furthermore, experimental observations or statistical inference enable us to better deduce the future possibilities from this additional acquired information about the dynamical evolution of state from the information entropy. \cite{PhysRev.106.620,PhysRev.108.171,mitchell_statistical_1969}

In these modern deep neural network topologies to predict the behavior of physical phenomenon, we utilized the physical laws which govern the time-dependent dynamics of a system, as ``prior knowledge'' and encode information as a regularization agent into the learning algorithm. The permissible phase-space of underline mechanics constrains these hybrid physical-centric informational networks. The generalized optimal solution of a feasible size is guided to inferred by rapidly augmenting the information from the available inadequate training validation dataset. State-of-the-art random forests, support vector machines, Gaussian processes regression, feed-forward recurrent and convolutional deep neural networks topology as a ``Black-box'' tool in the machine learning field give an excellent prediction in the abundant big-data regime of the image recognition, image segmentation, natural language task in the computer science and information technology domain. However, engineering systems and physical science are deficient data regimes, and extensive data acquisition is quite expansive. Conventional neural networks worsen in this inadequate data and partial information regime to predict and draw robust conclusions and make guaranteed decisions based on minimum uncertainty quantification and convergence. Recently, deeply learned artificial neural networks exceptionally over-perform humans when dealing with high-dimensional data set such as image classification and pattern segmentation, \cite{Krizhevsky_imagenetclassification_2012}, reinforcement learning, \cite{Yosinski11evolvingrobot,mnih2013playing,brockman2016openai,silver2017mastering,Schrittwieser_2020} natural language translation, \cite{Sutskever_Sequence_2014} robotic dexterity, \cite{openai2019learning,levine2016learning} and a specific type of strategic game playing. \cite{silver_mastering_2017} One reason for their outperforming humans is that neural networks operation harnesses the ``deep prior knowledge'' through translation invariances scheme embedded in their architecture through convolutional layers, self-attention layers, and batch normalization process to search for an optimal strategy in a well-conditioned energy landscape. However, despite the deep learning about their nearby environment, these machines still have flawed intuitions of nearby physical world kinematics and the dynamics operating principle, which governs the nearby environment. In comparison, our brain, the human neural networks from evolution, have this understanding, such as the abstraction of gravity while handling the Newtonian mechanics on earth in day-to-day work. For the machine to understand this ``feeling,'' needs a specialized network topology map to understand and sense it to incorporate in its prediction. \cite{Bakhtin_Phyre_2019} Physical sciences employed the underlying symmetries of nature to comprehend its functioning. Noether's theorem associates governing equations of physical science with these symmetry properties of nature through the notion of  ``a physics-prior'' of conservation laws of mass, charge, linear or angular momentum, spin, and energy. \cite{noether_invariante_1918,Hill_Hamilton_1951,Noether_Invariant} With the advancement of data science and deep neural network approaches in the recent past, there is a reinvigorated push to understand or reinterpret nature governing principles phenomenologically through a data informatics perspective. However, the new approach has promising results where abundant data is available with a specific statistical characteristic profile of the dataset to train the machine and predict the dynamics of nature. \cite{gibbs_elementary_1902,Gibbs_Thermodynamics_1961} However, deep neural network success in physical sciences interpretation is still modest because the existing neural network architecture and topology struggle to learn from data containing nature symmetries properties and conservation laws. In this regard, the physical mechanic has recently inspired the development of a novel network topology that can efficiently learn arbitrary conservation laws of nature. The objective is to directly associate the network with underlying symmetries through the ordinary differential equations (ODE) and partial differential equations (PDE) modeled for governing physical dynamics. \cite{Lagaris_Artificial_1998,Chen_Neural_2018,Karumuri_Simulator} Such a class of physics-inspired networks reported in the literature exploiting the generalized Hamiltonian mechanics, \cite{dirac_1950,Greydanus_Hamiltonian_2019,Toth_Hamiltonian_2020,finzi_simplifying_2020,Raissi_Physics_2019} and Lagrangian mechanics to train, learn and predict from the data that originated from following these physical dynamics by endowing neural networks with a prior physical. The Hamiltonian neural network topology is inspired by learning to conserve a quantity analogous to the total energy. The objective of the approach is to ameliorate the neural networks which do not have prior knowledge of the physical environment and learn in forwarding propagation from the dataset approximately about the physics laws instead of exact physical laws encoded by the generalized partial differential equation. This approximate physics laws information later minimizes in the backpropagation path as loss or cost function over the trained neural network through the stochastic statistical inference of physical knowledge.

\section*{\texorpdfstring{Lagrangian Mechanics}{Lagrangian Mechanics}}

We have epistemologically defined a dynamical evolution of the quantum state in a scalar Lagrangian density instead of a Hamiltonian approach. The Lagrangian mechanics is more general and does not constrain the canonical parameters in configuration space. We will use this Lagrangian density to train and map on a neural network and propose the Lagrangian density space-time neural network. Hartree product type time-dependent trial wave-functions are used to train the proposed neural network. We have used the Principle of Least or Stationary Action on this Lagrangian density, which evolves as second-order derivatives of the wave-function, giving us the approximate solution of Schr{\"o}dinger's \textit{et al.} \textit{Ansatz} of the dynamic of a quantum system. \cite{Schrodinger_Quantisierung1_1926,Schrodinger_Quantisierung2_1926}

In this regard, A Lagrangian density is defined by any generalized coordinate as a function of time and other derivatives of the generalized coordinates. The dynamical evolution of the non-equilibrium statistical ensemble prescribes that this Lagrangian density will take the most probable trajectory by maximizing the path information entropy, also known as the maximum caliber principle. Furthermore, the rate of increase of the information entropy can be phenomenologically understood as an act of potential physical field. On this potential by applying the variational approach yield ``principle of least dissipation of energy''. \cite{jaynes_minimum_1980,presse_principles_2013,hazoglou_communication_2015,davis_hamiltonian_2015} It is analogous to the probability amplitude assigned to a trajectory in the Feynman path integral formalism in quantum mechanics. \cite{feynman_theory_1963} 
Feynman's path integral approach is also based upon a classical particle Lagrangian to obtain a particle propagator. However, involved Lagrangian density is complex and contains second-order spatial and time derivatives. \cite{Feynman_Space-Time_1948} It was called ``seemingly artificial density''. Nevertheless, later proven through the variational principle application, it provides the direct solution of complex wave-function. \cite{schweber_introduction_1961,moiseiwitsch_variational_1966,lanczos_variational_1970,feynman_quantum_1965,nesbet_Variational_2002} After deducing the Lagrangian density, we know most about the quantum state. We can evaluate the path-integral kernel, continuity equation, conservation law of system constitute, response, and dissipation function behavior of the state with its surrounding environment. Moreover, this information-theoretical perspective of the dynamical evolution of a system is equivalent to the second law of thermodynamics for a physical system. \cite{hebb2005organization} Ontologically, a Hamiltonian density obtains by constraining the configuration space of Lagrangian density. The canonical procedure obtains canonical momenta is tensor conjugate with a generalized position coordinate. The canonical conjugate momenta and generalized coordinates satisfy the Hamiltonian operator's canonical commutation relation. This Hamilton's equation procedure uses to get energy Hamiltonian operator terms in the Schr{\"o}dinger equation. 
From the Lagrangian density next, we derive the continuity equation from the probability invariance of Lagrangian density in configuration space for the transport calculation. This Hamilton's Principle-based method gives an approximate solution of the time-dependent many-body system.

Recently ``Deep Lagrangian Networks'' (DeLaNs) have been built to continuously teach Lagrangians to control rigid body dynamics. \cite{lutter_deep_2018,cranmer_lagrangian_2020,gupta_a_2019,qin_machine_2020}  Instead of solving the closed-form analytical equation of governing mechanics, the central idea is to parameterize the arbitrary Hamiltonian and Lagrangian on the neural networks in terms of loss or cost function and search for the optimal solution. Lately, molecular Schr{\"o}dinger equation,\cite{Rupp_Fast_2012,ferminet} quantum chemical calculation,  \cite{schutt_quantum-chemical_2017} rigid-body dynamics on a graph network, \cite{Battaglia_Interaction_2016} domain-specific model are introduced. The final selection of a network topology depends upon the physical phenomena we are interested in investigating in relevant mechanics. The Hamiltonians neural network inspires by minimizing the physical dynamics' energy in the canonical coordinates to compute the canonical momenta and energy. These canonical coordinates are bound by the Poisson bracket rules of Hamiltonian mechanics. In the quantum mechanics counterpart, the Poisson bracket replaces the relevant operator bracket operation of mechanics. The Lagrangian density neural network strategy is further generalized and appropriate, as it does not restrict by the canonical coordinates of mechanics. Learned energies evaluate from the principle of least action in the functional form. In comparison, ``Hamiltonian Neural Networks'' (HNN) required stringent Poisson bracket conditions on canonical momenta dynamics. On the other hand, Lagrangian density networks restrict kinetic energy to any arbitrary Lagrangians. Lagrangian density deep neural networks approach combines many desirable properties. It can model dynamical systems learn from differential equations and exact conservation laws in arbitrary coordinates with arbitrary Lagrangians. \cite{Hill_Hamilton_1951}

\begin{figure}[H]
\centering
\includegraphics[width=\textwidth]{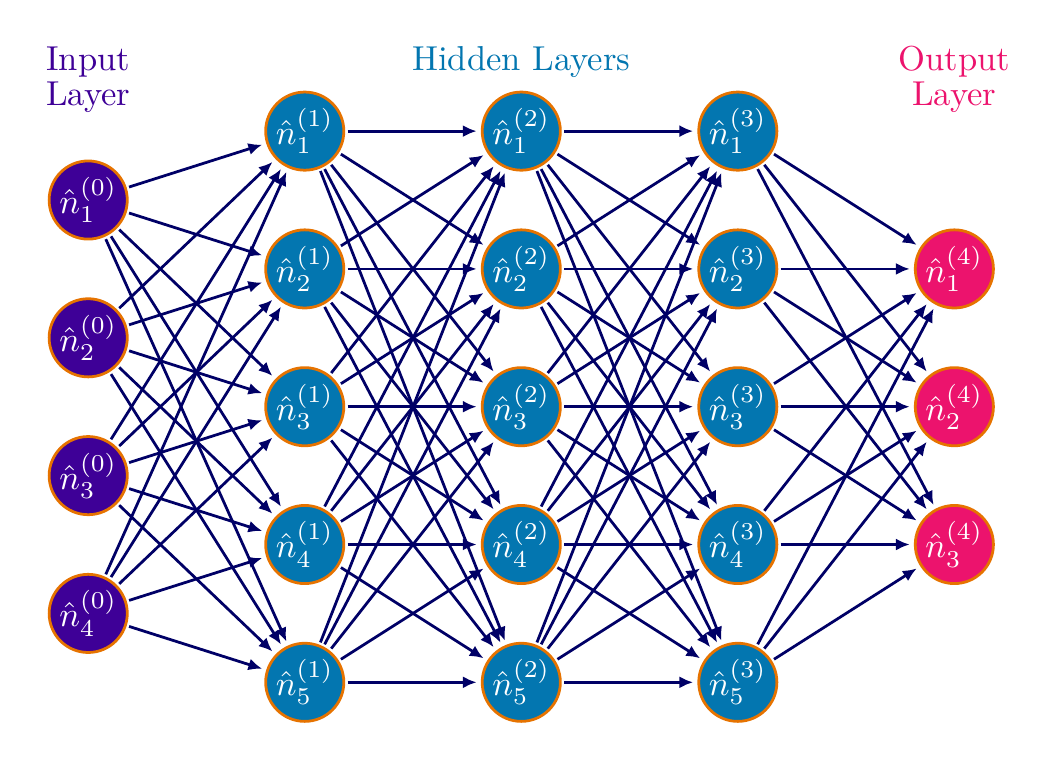}
\caption{Lagrangian density neural network}
\label{fig-0}
\end{figure}

\begin{figure}[H]
\centering
\includegraphics[width=\textwidth]{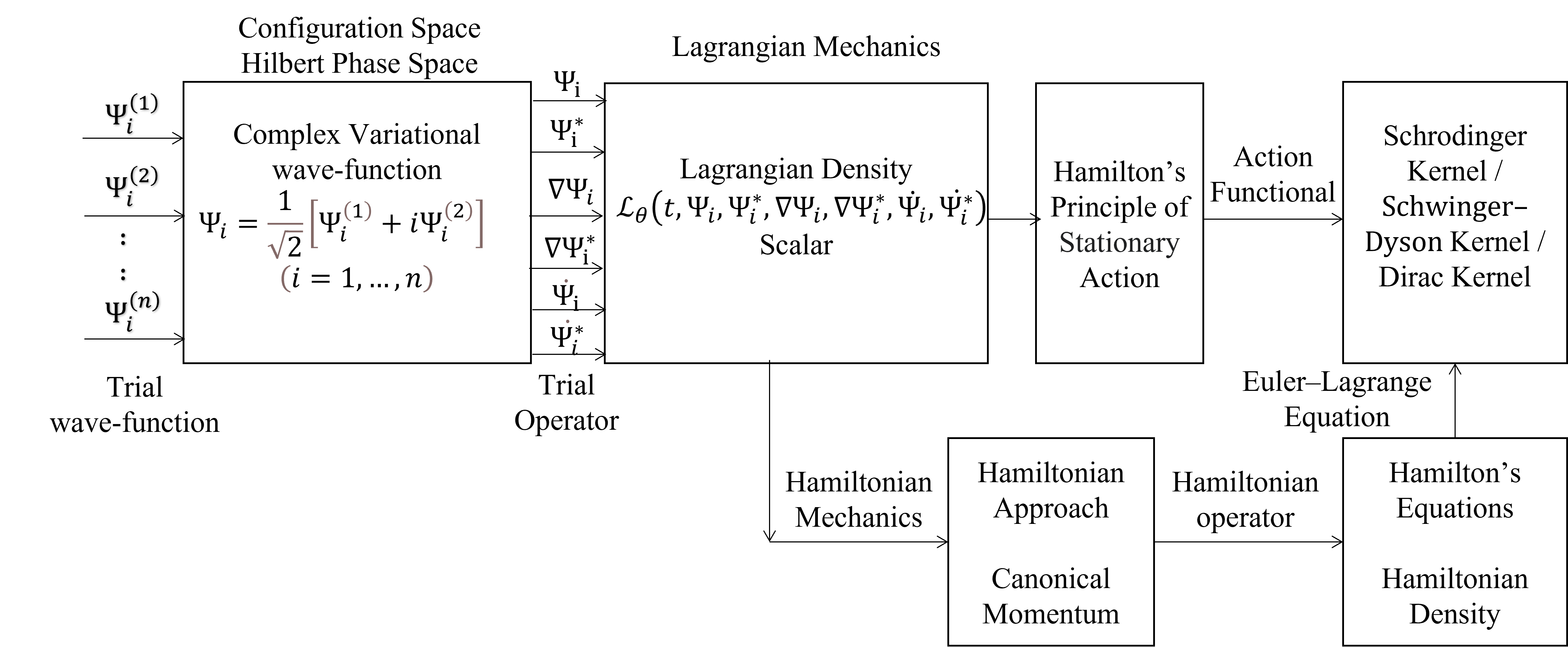}
\caption{Lagrangian density neural network flowchart}
\label{fig-1}
\end{figure}

\subsection*{\texorpdfstring{Classical Lagrangian Density}{Classical Lagrangian Density}}

In a generalized canonical position coordinates $ \boldsymbol{q}(t) $, and  generalized canonical momenta $ \boldsymbol{\dot q}(t) $ with time $ (t) $ connoting a Lagrangian density $ \mathscr{L}_{\theta}(\boldsymbol{q}(t),\boldsymbol{\dot q}(t),t) $ for a generic neural network's generalized learning parameter-set $ \theta $ as,

\begin{equation}\label{eq-1}
\begin{aligned}
\boldsymbol{\dot q}(t) \equiv \frac{\partial \mathscr{L}_{\theta}(\boldsymbol{q}(t),\boldsymbol{\dot q}(t),t)}{\partial \boldsymbol{\dot q}(t)}
\end{aligned}
\end{equation}
Where velocity vector $\boldsymbol{\dot q}(t) = \frac{d \boldsymbol{q}(t)}{d t} = \boldsymbol{v} $ is time derivative at $ \boldsymbol{q}(t) $, and by binary operation of Poisson bracket in Darboux coordinate systems of canonical position and momentum variables relation in phase-space and time also hold on their Poisson bracket as another defined
function in Hamiltonian mechanics $ \{\boldsymbol{q}_i,\boldsymbol{q}_j\}= 0 $, $ \{\boldsymbol{\dot q}_i,\boldsymbol{\dot q}_j\} = 0 $, $ \{\boldsymbol{q}_i,\boldsymbol{\dot q}_j\} = \delta_{ij} $ is Kronecker delta for $(i,j=1, \ldots, n )$ is defined as,

\begin{equation}\label{eq-2}
\{\mathcal {F},\mathcal {H}\}=\sum_{i}^{N}\bigg(\frac{\partial \mathcal {F}}{\partial \boldsymbol{q}_{i}} \frac{\partial \mathcal {H}}{\partial \boldsymbol{\dot q}_{i}}-\frac{\partial \mathcal {F}}{\partial \boldsymbol{\dot q}_{i}} \frac{\partial \mathcal {H}}{\partial \boldsymbol{q}_{i}}\bigg), \quad i=1, \ldots, n
\end{equation}

In the phase-space manifold, the trajectory time evolution $\mathcal{F}(\boldsymbol{q}(t),\boldsymbol{\dot q}(t), t)$ is,

\begin{equation}\label{eq-2-1}
\frac{d}{dt}\mathcal{F}(\boldsymbol{q}(t),\boldsymbol{\dot q}(t), t)={\frac {\partial \mathcal{F}}{\partial \boldsymbol{q}}}{\frac {d\boldsymbol{q}}{dt}}+{\frac {\partial \mathcal{F}}{\partial \boldsymbol{\dot q}}}{\frac {d\boldsymbol{\dot q}}{dt}}+\frac{\partial \mathcal{F}}{\partial t}
\end{equation}

From the Hamilton's solutions,

\begin{equation}\label{eq-2-2}
\begin{aligned}
{ {\boldsymbol{\ddot q}(t)}}={\frac {\partial \mathcal {H}}{\partial \boldsymbol{q}(t)}}=\{\boldsymbol{\dot q}(t),\mathcal {H}\} \\
{ {\boldsymbol{\dot q}(t)}}=-{\frac {\partial \mathcal {H}}{\partial \boldsymbol{\dot q}(t)}}=\{\boldsymbol{q}(t),\mathcal {H}\}
\end{aligned}
\end{equation}

Where statistical ensemble $\mathcal{F}(\boldsymbol{q}(t),\boldsymbol{\dot q}(t))$ is a distribution function of an observable in position-momenta phase space $ (\boldsymbol{q}(t),\boldsymbol{\dot q}(t)) $ and $\mathcal{H}$ is system Hamiltonian energy. In terms of Poisson bracket Hamilton's equations of motion is,

\begin{equation}\label{eq-2-3}
\begin{aligned}
\frac{d}{dt}\mathcal{F}(\boldsymbol{q}(t),\boldsymbol{\dot q}(t), t)&=\bigg({\frac {\partial \mathcal{F}}{\partial \boldsymbol{q}}}{\frac {\partial \mathcal {H}}{\partial \boldsymbol{\dot q}}}-{\frac {\partial \mathcal{F}}{\partial \boldsymbol{\dot q}}}{\frac {\partial \mathcal {H}}{\partial \boldsymbol{q}}}\bigg)+{\frac {\partial \mathcal{F}}{\partial t}}=\{\mathcal {F},\mathcal {H}\}+{\frac {\partial \mathcal {F}}{\partial t}}
\end{aligned}
\end{equation}

In this phase-space manifold a probability $\rho (\boldsymbol{q}(t),\boldsymbol{\dot q}(t)) $ distribution in the infinitesimal phase-space volume
$  {d} ^{n}\boldsymbol{q}\,\mathrm {d} ^{n}\boldsymbol{\dot q}$ govern through the Liouville equation that demonstrates the conservation of density in phase space.

\begin{equation}\label{eq-2-4}
\frac{d\rho (\boldsymbol{q}(t),\boldsymbol{\dot q}(t))}{dt}=
\frac{\partial\rho}{\partial t}
+\sum_{i=1}^N\left(\frac{\partial\rho}{\partial \boldsymbol{q}_i}\dot{\boldsymbol{q}}_i
+\frac{\partial\rho}{\partial \boldsymbol{\dot q}_i}{\boldsymbol{\ddot q}}_i\right)=0
\end{equation}

And, in Poisson bracket terms the Liouvillian density is, 

\begin{equation}\label{eq-2-5}
\frac{\partial \rho}{\partial t}=-\{\rho,\mathcal{H}\}
\end{equation}

Where, again $\mathcal{H}$ is system Hamiltonian energy.

\subsection*{\texorpdfstring{Quantum Counterpart}{Quantum Counterpart}}

Upon quantization in generalized Lie algebra, Poisson brackets deform into Moyal brackets,\cite{moyal_1949}, and from Dirac's thesis \textit{et al.},\cite{Dirac_1925} Poisson brackets supplanted by commutators operator of the quantized system by mapping a classical phase-space function $\rho$, and $ \mathcal{H} $ on the state-space density operator $ {\widehat{\rho}} $ and Hamiltonian operator $ \widehat{\mathcal{H}} $ on the complex Hilbert space as,

\begin{equation}\label{eq-2-6}
\{\rho, \mathcal{H}\} \longmapsto \frac{1}{i \hbar}[\widehat{\rho}, \widehat{\mathcal{H}}]
\end{equation}

And, the density operator time evolution in the  Schr{\"o}dinger representation as a mixed state by Liouvillevon-Neumann equation is, 

\begin{equation}\label{eq-2-7}
\frac{\partial \rho}{\partial t}=\frac{1}{i\hbar}[ {\mathcal{H}},{\rho}]
\end{equation}

Where $ \rho $ is the density matrix , and square brackets denote a commutator operation.

However, in the Heisenberg representation, where Hamiltonian operator $ \widehat{\mathcal{H}} $ is time-dependent while the wave-functions are now time-independent, the Heisenberg equation of motion has a crucial sign difference from the \cref{eq-2-7} as,

\begin{equation}\label{eq-2-8}
\frac{d }{d t}\langle\widehat{\Psi}_{H}\rangle=-\frac{1}{i\hbar}[\widehat{\mathcal{H}}, \widehat{\Psi}_{H}]
\end{equation}

Where $ \widehat{\Psi}_{H} $ is time-dependent Heisenberg operator of Schr{\"o}dinger wave-function $ {\Psi}(\boldsymbol{r}, t) $ and time derivative of the expected value $ \langle\widehat{\Psi}_{H}\rangle $ is same due to relative negative sign.

Corresponding equation in Schr{\"o}dinger representation by Ehrenfest's theorem is, 

\begin{equation}\label{eq-2-9}
\frac{d}{d t}\langle \widehat{\Psi}(\boldsymbol{r}, t)\rangle=\frac{1}{i \hbar}\langle[\widehat{\Psi}, \widehat{\mathcal{H}}]\rangle
\end{equation}

Where state is time-dependent and operator
is stationary. A general relation between the expectation of a quantum mechanical density operator $ \widehat{\rho} $ and the expectation of the commutator of that operator with the Hamiltonian of the system is, 

\begin{equation}\label{eq-2-10}
\frac{d}{d t}\langle \rho\rangle=\frac{1}{i \hbar}\langle[\widehat{\rho}, \widehat{\mathcal{H}}]\rangle+\langle\frac{\partial \widehat{\rho}}{\partial t}\rangle
\end{equation}

where $ \widehat{\rho}$ is  quantum mechanical density operator and $ \langle {\rho} \rangle $ is density expectation value. Mathematically, $ { {\widehat {\rho }}} $ is a positive-semidefinite Hermitian matrix with unit trace. Historically, Lev Landau introduced the density matrix because a subsystem of a composite quantum system was impossible to describe by a state vector. However, Von Neumann introduces the density matrix treatment to develop a theory of quantum measurements and quantum statistical mechanics. 

\subsection*{\texorpdfstring{Lagrangian Denisty Learning Topology}{Lagrangian Denisty Learning Topology}}

We will learn from the Lagrangian denisty in the arbitrary coordinates system enforcing conservation of total energy. In Lagrangian mechanics, a system in pair of $(\mathcal{M}, \mathscr{L}_{\theta})$ in configuration space $(\mathcal{M})$ where all ${\boldsymbol{q}_{i,j,k}\in \mathcal{M}}$ and smooth function $ \mathscr{L}_{\theta}=\mathscr{L}_{\theta}(\boldsymbol{q}(t),\boldsymbol{\dot q}(t),t) $ with a kinetic energy $ {T} $ and potential energy $ {V} $, the Lagrangian density describing the energies dynamics of the entire system is $ {\mathscr{L}_{\theta}\equiv T-V} $. The time evolution of the system in a coordinates $x_t = (\boldsymbol{q}, \boldsymbol{\dot q})$ from starting state of $ {x_{0}:[t_{0}]\to \mathcal{M}} $, to the end state of $ {x_{1}:[t_{1}]\to \mathcal{M}} $, can take many trajectories in the configuration space $(\mathcal{M})$, from the Lagrangian mechanics, the  stationary point in this evolution is defined as ``the action'' functional,

\begin{equation}\label{eq-3}
{{\mathcal{S}}[\boldsymbol{q}]\,{\stackrel {\text{def}}{=}}\,\int _{t_{0}}^{t_{1}}\mathscr{L}_{\theta}(\boldsymbol{q}(t),{\boldsymbol{\dot q}}(t),t) dt}\equiv \int_{t_0}^{t_1} [T(\boldsymbol{q}(t), \boldsymbol{\dot q}(t)) - V(\boldsymbol{q}(t))]dt
\end{equation}

The configuration space $(\mathcal{M})$ of ${\mathbb {C} ^{n}}$ in Hilbert space of a Cauchy complete metric space $M$, starting state time instants $t_{0}$ to final state time instants $t_{1}$, a smooth path $ \boldsymbol{q}_{0}:[t_{0}, t_{1}]\to \mathcal{M} $ is a stationary point of $\mathcal{S}[\boldsymbol{q}]$, if all its directional derivatives at $\boldsymbol{q}_{0} $ vanish, i.e., for every smooth $ {\delta :[t_{0},t_{1}]\to \mathbb {C} ^{n},}$ which state that system will take the path for which $\mathcal{S}[\boldsymbol{q}]$ is stationary $\delta \mathcal{S}[\boldsymbol{q}]=0$ minimum value from the principle of stationary action as,

\begin{equation}\label{eq-3-1}
{\delta\mathcal{S}[\boldsymbol{q}]\ {\stackrel {\text{def}}{=}}\ {\frac {d}{d\varepsilon }}{\Biggl |}_{\varepsilon =0}{\mathcal{S}}\left[\boldsymbol{q}_{0}+\varepsilon \delta \right]=0}
\end{equation}

Where directional derivative $\delta\mathcal{S}[\boldsymbol{q}]$ is defined as variation in physics and Gateaux derivative in mathematics, and function $ \delta (t) $ is perturbation or virtual displacement. The Lagrangian density neural network learned from the action functional and predicted the energy optima solution that gives system evolution's stationary state. From the constrain of \cref{eq-3} $\mathscr{L}_{\theta}\equiv T-V $ on the system, the path of the system described by the Euler-Lagrange equation from the calculus of variations is,

\begin{equation}\label{eq-4}
\frac{d}{dt} \frac{\partial \mathscr{L}_{\theta}(\boldsymbol{q}(t),{\boldsymbol{\dot q}}(t),t)}{\partial \boldsymbol{\dot q}_i} - \frac{\partial \mathscr{L}_{\theta}(\boldsymbol{q}(t),{\boldsymbol{\dot q}}(t),t)}{\partial\boldsymbol{q}_i} = 0,  \quad i=1, \ldots, n
\end{equation}

To solve the time evolution of the physical  system in science, we solve for the analytical expression of Lagrangian density $ \mathscr{L}(\boldsymbol{q}(t),\boldsymbol{\dot q}(t),t) $ in the Euler-Lagrange differential \cref{eq-4} equation. In the data-informatics solution scheme through the deep neural networks, we will first parametrize the Lagrangian density $ \mathscr{L}_{\theta}(\boldsymbol{q}(t),\boldsymbol{\dot q}(t),t) $ as a black box with a generic neural network's generalized learning parameter-set $ \theta $. Further, we will begin to deduce the dynamics from the trained network. To parametrize the network, we will first vectorize the \cref{eq-4} and then derive a loss or cost function, which in the backpropagation stage we try to optimize as follow,

\begin{equation}\label{eq-5}
\nabla_{\boldsymbol{q}} \mathscr{L}_{\theta}(\boldsymbol{q}(t),{\boldsymbol{\dot q}}(t),t) = \frac{d}{dt} \nabla_{\boldsymbol{\dot q}} \mathscr{L}_{\theta}(\boldsymbol{q}(t),{\boldsymbol{\dot q}}(t),t) 
\end{equation}

Where operator $[\nabla_{\boldsymbol{\dot q}}]_i\equiv \frac{\partial}{\partial {\boldsymbol{\dot q}}_i}$ are matrices. From the product and chain rules properties of a functional derivative, the time derivative of a symplectic gradient of Lagrangian density expanded in terms of $\boldsymbol{\dot q}$ and $\boldsymbol{\ddot q}$ as,

\begin{equation}\label{eq-6}
\nabla_{\boldsymbol{q}} \mathscr{L}_{\theta}(\boldsymbol{q}(t),{\boldsymbol{\dot q}}(t),t) = \big[\nabla_{\boldsymbol{q}}\nabla_{\boldsymbol{\dot q}}^{\intercal}\mathscr{L}_{\theta}(\boldsymbol{q}(t),{\boldsymbol{\dot q}}(t),t)\big] \boldsymbol{\dot q} + \big[\nabla_{\boldsymbol{\dot q}}\nabla_{\boldsymbol{\dot q}}^{\intercal}\mathscr{L}_{\theta}(\boldsymbol{q}(t),{\boldsymbol{\dot q}}(t),t)\big]\boldsymbol{\ddot q} 
\end{equation}

Where operator products $[\nabla_{\boldsymbol{q}}\nabla_{\boldsymbol{\dot q}}^{\intercal}\mathscr{L}_{\theta}(\boldsymbol{q}(t),{\boldsymbol{\dot q}}(t),t)]_{ij} = \frac{\partial^2 \mathscr{L}_{\theta}(\boldsymbol{q}(t),{\boldsymbol{\dot q}}(t),t)}{\partial \boldsymbol{\dot q}_i \partial \boldsymbol{q}_j }$ are also matrices. Using the efficient matrix inversion operation $\mathcal{O}(d)$, in comparison conventional green function calculation, matrix inversion operation expands with order 3 of dimension $\mathcal{O}(d^3)$. The generalized momenta $\boldsymbol{\ddot q}$ of the system is,

\begin{equation}\label{eq-7}
\boldsymbol{\ddot q} = \bigg[\nabla_{\boldsymbol{\dot q}}\nabla_{\boldsymbol{\dot q}}^{\intercal}\mathscr{L}_{\theta}(\boldsymbol{q}(t),{\boldsymbol{\dot q}}(t),t)\bigg]^{-1}\bigg[\nabla_{\boldsymbol{q}} \mathscr{L}_{\theta}(\boldsymbol{q}(t),{\boldsymbol{\dot q}}(t),t) - \big[\nabla_{\boldsymbol{q}}\nabla_{\boldsymbol{\dot q}}^{\intercal}\mathscr{L}_{\theta}(\boldsymbol{q}(t),{\boldsymbol{\dot q}}(t),t)\big] \boldsymbol{\dot q}\bigg]
\end{equation}

Where Hessian matrices $ H_{\mathscr{L}_{\theta}} $ of Lagrangian $ {\mathscr{L}_{\theta}:\mathbb {C} ^{n}\to \mathbb {C} } $  is,
\begin{equation}\label{eq-7-1}
H_{\mathscr{L}_{\theta}}(\boldsymbol {q}(t) ,\boldsymbol {\dot {q}}(t) ,t)= [\nabla_{\boldsymbol{\dot q}}\nabla_{\boldsymbol{\dot q}}^{\intercal}\mathscr{L}_{\theta}(\boldsymbol{q}(t),{\boldsymbol{\dot q}}(t),t)]^{-1}
\end{equation}

And, from \cref{eq-7-1} into \cref{eq-7},

\begin{equation}\label{eq-7-2}
\boldsymbol{\ddot q} = H_{\mathscr{L}_{\theta}}(\boldsymbol {q}(t) ,\boldsymbol {\dot {q}}(t) ,t)\bigg[\nabla_{\boldsymbol{q}} \mathscr{L}_{\theta}(\boldsymbol{q}(t),{\boldsymbol{\dot q}}(t),t) - \big[\nabla_{\boldsymbol{q}}\nabla_{\boldsymbol{\dot q}}^{\intercal}\mathscr{L}_{\theta}(\boldsymbol{q}(t),{\boldsymbol{\dot q}}(t),t)\big] \boldsymbol{\dot q}\bigg]
\end{equation}

And, $ \big[\nabla_{\boldsymbol{q}}\nabla_{\boldsymbol{\dot q}}^{\intercal}\mathscr{L}_{\theta}(\boldsymbol{q}(t),{\boldsymbol{\dot q}}(t),t)\big] $ is first-order Jacobian matrix product of Lagrangian density.
The dynamical solution from the Lagrangian density black box, obtained by integrating the $\boldsymbol{\ddot q}(t) $ trajectories from the different set of coordinates $x_t=(\boldsymbol{q}(t),\boldsymbol{\dot q}(t))$ from starting state of $ {x_{0}:[t_{0}]\to \mathcal{M}} $, to the end state of $ {x_{1}:[t_{1}]\to \mathcal{M}} $, which take place in the configuration space $(\mathcal{M})$. The cost or loss function in terms of the difference between square root of the sum of the squared $\boldsymbol{\ddot q}(t)_{\mathscr{L}_{\theta}} $ and $\boldsymbol{\ddot q}(t)_{\textrm{true}} $ as 2-norm is,

\begin{equation}\label{eq-8}
\textrm{Lagrangian Density Loss Function} \;\; \mathcal{L}_{\mathscr{L}_{\theta}}=\|\boldsymbol{\ddot q}(t)_{\mathscr{L}_{\theta}}-\boldsymbol{\ddot q}(t)_{\textrm{true}}\|_{2}
\end{equation}

\subsection*{\texorpdfstring{Lagrangian Denisty Neural Network Implementation}{Lagrangian Denisty Neural Network Implementation}}

In the physical dynamics-informed neural network topology, the biases and weights on the nodes are maps that underline governing mechanism. The neural network learns invariant quantities by endowing the ``physical a prior'' through approximating the Lagrangian action. The Euler-Lagrange \cref{eq-7} solved through \texttt{JAX} framework from \texttt{Google}, \cite{jax2018github,47008}  which is a high-performance accelerator code including  \texttt{Python} and \texttt{Numpy} machine learning package. 
In the forward propagation path, Hessian matrices $ H_{\mathscr{L}_{\theta}}(\boldsymbol {q}(t) ,\boldsymbol {\dot {q}}(t) ,t)= [\nabla_{\boldsymbol{\dot q}}\nabla_{\boldsymbol{\dot q}}^{\intercal}\mathscr{L}_{\theta}(\boldsymbol{q}(t),{\boldsymbol{\dot q}}(t),t)]^{-1}$ inverted which scale with $\mathcal{O}(n^3)$ with $n$ is the number of coordinates. In the \texttt{JAX} it is implemented as assuming Lagrangian density $\mathscr{L}_{\theta}$ is a differentiable function with three input vectors and one scalar output. The velocity vector $ \boldsymbol{\dot q}(t) $ of  $ \boldsymbol{q}(q) $ is represented as \texttt{q\_t} and \texttt{q\_tt} represent $ \boldsymbol{\ddot q}(t) $ accelerations  of $ \boldsymbol{q}(q) $, and vector \texttt{m} is a non-dynamical parameters as,

\begin{verbatim}
q_tt = (jax.numpy.linalg.pinv(jax.hessian(lagrangian, 1)(q, q_t, m))
@ (jax.grad(lagrangian, 0)(q, q_t, m)- jax.jacobian(jax.jacobian(lagrangian, 1), 0)
(q, q_t, m) @ q_t))
\end{verbatim}

Loss function with parameters in the Lagrangian density network is called as \texttt{loss(params, \ldots)}, and the symplectic gradient is called through \texttt{jax.grad(loss, 0)(params, \ldots)}.

We implemented a four-layer network with 500 hidden units with a batch size of 32 and a learning rate starting at $10^{-3}$. In our network, we calculate the second-order derivative of Hessian matrices to calculate the Lagrangian density; therefore, we have to choose the activation function that is not zero in the second-order derivative. The commonly employed \texttt{ReLU} function is zero in a second-order derivative; hence \texttt{tanh}, \texttt{sigmoid}, and \texttt{softplus} activations should be used. We found \texttt{softplus} perform well in learning convergence.

Similar to the Hamiltonian graph network approach, \cite{Alvaro_Hamiltonian_2019,cranmer2020discovering} We can learn a Lagrangian density by summing over the total Lagrangian from the \cref{eq-7}. The density is estimated as in terms of independent probability amplitude wave displacement $\boldsymbol{\phi}_{i,t}$ at each gridpoint $x_{i,t} = (\boldsymbol{q}(t), \boldsymbol{\dot q}(t))$ and by taking the sum over all the connected $n$ gridpoint, the total Lagrangian density is,

\begin{equation}\label{eq-9}
\mathscr{L}_{\theta} = \sum_{i=1:n}\mathscr{L}_{\theta_i},
\text{ for }\mathscr{L}_{\theta_i} =\mathscr{L}_{\theta}[\{\boldsymbol{\phi}_j, \dot{\boldsymbol{\phi}}_j\}_{j\in{\mathcal{I}}_i} ], \quad i=1, \ldots, n
\end{equation}

Where $ \{\boldsymbol{\phi}_j, \dot{\boldsymbol{\phi}}_j\} $ is again Poisson bracket and  ${\mathcal{I}}_i=\{i, \ldots\}$ is the set of indices connected to $i^{th}$ gridpoint. For a 2-D lattice grid, it is the nearest neighbor $\{(i,j), (i\pm1,j\pm1)\}$ which influences the dynamics of the central gridpoint, and for a 1-D grid, adjacent gridpoints $\{i-1,i,i+1\}$. By \cref{eq-7} Lagrangian density in $\boldsymbol{\phi}$ coordinates system as,

\begin{equation}\label{eq-10}
\boldsymbol{\ddot \phi} = \bigg[\nabla_{\boldsymbol{\dot \phi}}\nabla_{\boldsymbol{\dot \phi}}^{\intercal}\mathscr{L}_{\theta}(\boldsymbol{\phi}(t),{\boldsymbol{\dot \phi}}(t),t)\bigg]^{-1}\bigg[\nabla_{\boldsymbol{\phi}} \mathscr{L}_{\theta}(\boldsymbol{\phi}(t),{\boldsymbol{\dot \phi}}(t),t) - \big[\nabla_{\boldsymbol{\phi}}\nabla_{\boldsymbol{\dot \phi}}^{\intercal}\mathscr{L}_{\theta}(\boldsymbol{\phi}(t),{\boldsymbol{\dot \phi}}(t),t)\big] \boldsymbol{\dot \phi}\bigg]
\end{equation}

Where, $\nabla_{\boldsymbol{\phi}} \equiv \{\frac{\partial}{\partial\boldsymbol{\phi}_1}, \frac{\partial}{\partial\boldsymbol{\phi}_2}, \ldots, \frac{\partial}{\partial\boldsymbol{\phi}_n}\}$. By selecting the nearest neighbor grid point, the Hessian matrix in \cref{eq-10} became sparse with non-zero entries for all next-nearest neighbor grid points. Therefore, it becomes much more efficient to invert in back-propagation cycle of Lagrangian density auto-gradient operation. In \texttt{JAX}, it can be easily vectorized for forward and back-propagation over the grid. The Lagrangian density is modeled as a multi-layer perceptron graph neural network. We can understand it intuitively instead of the regular Lagrangian network where we calculate Lagrangian from a fixed set of coordinates similar to the hamiltonian network scheme. Here we accumulate Lagrangian density on the coordinates connected in the nearest neighbor configuration, and total density spread over these points and for a different configuration adjacency matrix change by $\mathcal{I}_i$ on the graph.

The Lagrangian density approach has two-fold advantages compared to the mapping differential equation of the system on the ordinary neural network. First, Lagrangian density is scalar functional and easier to learn by the neural network than the system's dynamical differential operator. Second, integrating the differential gradient from learned Lagrangian density predicts approximately the dynamical trajectory solution conserving the underline law of physics. Furthermore, High-dimensional elliptic and parabolic partial differential equations approximate solution through numerical technique suffer from the curse of dimensionality and the number of mathematical operations exploding in computer memory storage and time. However, the approximate solution of these stochastic partial differential equations through the deep neural network does not suffer from the curse of dimensionality. It grows at most polynomially in the space-time region. \cite{Robbins_Stochastic_1951,hornung2020spacetime,nusken2020solving}

Lagrangian density-informed neural network topology's methodological and algorithmic development is an endowing promising approach for assorted physical science and engineering system dynamical solutions in the data-driven mathematical forecasting, multi-scale, multi-physics scientific computational modeling, and simulation. In the future, we hope these new physics-informed deep neural networks will coexist in a fruitful synergy as a co-solver with the already matured conventional finite element method, fourth-order Runge–Kutta integrator, spectral-based numerical solvers. Furthermore, provide invaluable intuition in constructing structured solutions of physical systems, potentially enriching the scientific computation field.

The predictive modeling and algorithm's growth from data deprives sway of a physical system is a challenging task. To design the network and its optimal model, we must consider broad investigation topics from information science and physical science. To increase the prediction accuracy, we have to investigate how the associated data uncertainty propagated through the network and its quantification. 
We must consider the efficient dataset size for the network, the quality of data scattered across its input parameters, and the depth and width selection of the network layers for the corresponding dataset. Are the algorithm suffering and sticking with local optima for the dataset. Are the differential operators of the physical system converging to unique values, and are they suffering from vanishing gradients in the backpropagation in the hidden neural network and higher-order differential operators. The deep neural networks also suffer from spectral bias during the learning phase. \cite{pmlr-v97-rahaman19a} Therefore if it is trained first on the low-frequency components feed of parameters, it struggles to learn later on the high-frequency components and forever tries to learn to these components and never converges. Hence the learning order of the dataset should be carefully selected. However, it can be overcome by Fourier feature mapping to the input dataset, where we pass all the neural network inputs through an added Fourier component. Hence, the neural network also learns and achieves faster convergence for high-frequency components. \cite{tancik2020fourfeat} We should also consider the search for optimal schemes for initializing weights and inductive biases of the hidden layers and activation function choices based on the underlying physical dynamics equation. Defining the appropriate loss/cost functions infer the system's physical dynamics. The system-specific tunning of the network will enhance results for distinct quality of differential equations governing the physical system. Admittedly, two entirely different internal neural network structures and topology may give rise to a similar optimal prediction through the way information is inference in the network. Hence, it is always necessary to ask before finalizing and proposing a deep learning solution whether the nearby optimal solution or other network topology is available, which is computational and energy-efficient, to give similar robust predication. These are some of the current challenges in solving partial and ordinary differential equation from a data science perspective.

\subsection*{\texorpdfstring{Quantum-State Lagrangian Denisty}{Quantum-State Lagrangian Denisty}}

The non-relativistic Schr{\"o}dinger wave-equation is, 

\begin{equation}\label{eq-15}
i \hbar \frac{\partial}{\partial t} \Psi(\boldsymbol{r}, t)=-\frac{\hbar^{2}}{2 m} \frac{\partial^{2}}{\partial \boldsymbol{r}^{2}}\Psi(\boldsymbol{r}, t) +V(\boldsymbol{r}, t)\Psi(\boldsymbol{r}, t) 
\end{equation}

The Lagrangian density for a Schr{\"o}dinger wave from the argument of \cref{eq-1} is,

\begin{equation}\label{eq-16}
\mathscr{L}_{\theta}=\frac{\hbar^{2}}{2 m}[\frac{\partial}{\partial \boldsymbol{r}} \Psi(\boldsymbol{r}, t) \cdot\frac{\partial}{\partial \boldsymbol{r}} \Psi^{\dagger}(\boldsymbol{r}, t)]+V \Psi(\boldsymbol{r}, t) \Psi^{\dagger}(\boldsymbol{r}, t)+i \hbar[\Psi(\boldsymbol{r}, t) \frac{\partial}{\partial t} \Psi^{\dagger}(\boldsymbol{r}, t)-\Psi^{\dagger}(\boldsymbol{r}, t) \frac{\partial}{\partial t} \Psi(\boldsymbol{r}, t)]
\end{equation}

A quantum system define by variational wave-functions $\Psi_{i}(\boldsymbol{r},t)$, where (i=1, \ldots, n) $n$ components complex variational parameters and $ \boldsymbol{r}= \sum_{j=1}^{n} \boldsymbol{r}_j = (\boldsymbol{x,y,z}) = (\sigma_{\boldsymbol{x}},\sigma_{\boldsymbol{y}},\sigma_{\boldsymbol{z}})$ is generalized space-spin variables coordinates in Hilbert space with time $ (t)$.

\begin{equation}\label{eq-A1}
\Psi_{i}=\frac{1}{\sqrt{2}}\big[\Psi_{i}^{(1)}+i \Psi_{i}^{(2)}\big] \quad(i=1, \ldots, n)
\end{equation}

Where $\Psi_{i}^{(1)}$ and $\Psi_{i}^{(2)}$ are real functions. Assuming that $\Psi_{i}^{(1)}$ and $\Psi_{i}^{(2)}$ are all entirely independent of each other.
$\Psi_{i}(\boldsymbol{r},t)$ can be constructed by time-dependent Hartree-Fock product type identical, normalized time-dependent single-particle trial wave-function $\phi(\boldsymbol{r},t)$ with Slater determinant as,

\begin{equation}\label{eq-A2}
\Psi(\boldsymbol{r}_{1},\boldsymbol{r}_{2},\ldots,\boldsymbol{r}_{n}, t)\equiv\ket{\phi(\boldsymbol{r}_{1},t)\phi(\boldsymbol{r}_{2},t)\ldots \phi(\boldsymbol{r}_{n},t)} 
\end{equation}

The Lagrangian density of a quantum state defined of wave-function and its complex conjugate, partial spatial and temporal derivatives of the wave-function and its complex conjugate, time of up to any arbitrary order as,

\begin{equation}\label{eq-17}
\mathscr{L}_{\theta} \Big(t, \Psi_{i}(\boldsymbol{r},t),\Psi_{i}^{*}(\boldsymbol{r},t),\frac{\partial \Psi_{i}(\boldsymbol{r},t)}{\partial \boldsymbol{r}}, \frac{\partial \Psi_{i}^{*}(\boldsymbol{r},t)}{\partial \boldsymbol{r}}, \frac{\partial \Psi_{i}(\boldsymbol{r},t)}{\partial t},\frac{\partial \Psi_{i}^{*}(\boldsymbol{r},t)}{\partial t}\Big) = \mathscr{L}_{\theta}\Big(t, \Psi_{i}, \Psi_{i}^{*}, \nabla \Psi_{i}, \nabla \Psi_{i}^{*}, \dot\Psi_{i}, \dot\Psi_{i}^{*}\Big)
\end{equation}

Which is a function of complex wave-functions $\Psi_{i}(\boldsymbol{r},t)$ and it's complex conjugate $\Psi_{i}^{*}(\boldsymbol{r},t)$ components, their spatial derivatives $\nabla \Psi_{i} = \partial \Psi_{i} / \partial \boldsymbol{r}$, time derivatives $\dot\Psi_{i} = \partial \Psi_{i} / \partial t$ and time coordinates $t$. 

\subsection*{\texorpdfstring{Lagrangian Denisty Action Functional}{Lagrangian Denisty Action Functional}}

The partial spatial and temporal derivatives of wave-functions expanded to any arbitrary higher-order give the equivalent Lagrangian density. It gives the same action functional upon integration over hyper-volume as defined ahead. Here, we will continue the derivation with the first-order partial derivative. Independent variations of the functions $\Psi_{i}(\boldsymbol{r},t)$ and $\Psi_{i}^{*}(\boldsymbol{r},t)$ lead to two complex conjugate identity. Now in the $\boldsymbol{r}$ dimensional space, define an Lagrangian density integral $ I(L) $ for $\Psi_{i}(\boldsymbol{r},t)$ whose domain of integration is in hyper-volume $V$, in the time interval $ t_{1}$ to $t_{2} $ lead to action functional $\mathcal {S}[\Psi^{*},\Psi]$ as,

\begin{equation}\label{eq-18}
\mathcal {S}[\Psi^{*}, \Psi]= I(L)=\int_{T} L dt =\int_{t_{1}}^{t_{2}} dt\int_{V} \mathscr{L}_{\theta}\Big(t, \Psi_{i}(\boldsymbol{r},t), \frac{\partial \Psi_{i}(\boldsymbol{r},t)}{\partial \boldsymbol{r}}, \frac{\partial \Psi_{i}(\boldsymbol{r},t)}{\partial t}\Big)d\boldsymbol{r} =\int_{T} dt \int \mathscr{L}_{\theta}\Big(t, \Psi_{i}, \nabla \Psi_{i}, \dot\Psi_{i}\Big)d^{3n} \boldsymbol{r}
\end{equation}

Where the integration of Lagrangian function $ L $  is insofar over total time between $ t_{1}$ to $t_{2} $ and entire configuration space with an element of volume $d^{3n}\boldsymbol{r}=d^{3}\boldsymbol{r}_{1}d^{3}\boldsymbol{r}_{2}\ldots d^{3}\boldsymbol{r}_{n}$, and returns a scalar value. The dimensions of action in physical interpretation are the same as angular momentum, energy-time or Plank constant, and momentum-length. In information science, it can be interpreted as quantification of uncertainty or subjective probability as how much nature allows us to know about the degree of freedoms of its hidden variables. 

The variation of the wave-functions components defines as,

\begin{equation}\label{eq-20}
\Psi_{i}(\boldsymbol{r},t) \rightarrow \Psi_{i}(\boldsymbol{r},t)+ \delta \Psi_{i}(\boldsymbol{r}, t) \equiv \Psi_{i}(\boldsymbol{r},t)+\epsilon \eta_{i}(\boldsymbol{r},t)
\end{equation}

Where the $\eta_{i}(\boldsymbol{r},t)$ are any set of linearly independent functions of $(\boldsymbol{r},t)$ which vanish over the hyper-surface $S$ boundary in the integration region of $V$ volume satisfying the boundary condition. 
The variation of wave-functions components defines as,

\begin{equation}\label{eq-21}
\delta \Psi_{i}(\boldsymbol{r}, t_{1})=\delta \Psi_{i}(\boldsymbol{r}, t_{2})=0
\end{equation}

The stationary solution of dynamical equations through the Lagrangian density integral $ I(L) $ obtained from Hamilton's Principle of Least or stationary Action. \cite{Hill_Hamilton_1951}
From the variational principle method, the action functional variation $\delta \mathcal {S}[\Psi^{*},\Psi]$ for the Lagrangian density $\mathscr{L}_{\theta}$ is, \cite{PhysRev.152.1219,Zwanzig_Time_1965}

\begin{equation}\label{eq-19}
\delta \mathcal {S}[\Psi^{*}, \Psi] = \delta I(L)= \int_{t_{1}}^{t_{2}} \delta L \;\; dt = \int_{T} dt \int \delta\mathscr{L}_{\theta}(\Psi,\Psi^{*},\ldots)d^{3n} \boldsymbol{r}=0\rightarrow\text{stationary}
\end{equation}

Where the variation of Lagrangian $ \delta L $ by definition is,

\begin{equation}\label{eq-19-1}
\delta L=\sum_{j=1}^{n}\bigg(\frac{\partial L}{\partial q_{j}} \delta q_{j}+\frac{\partial L}{\partial \dot{q}_{j}} \delta \dot{q}_{j}\bigg), \quad \delta \dot{q}_{j} \equiv \delta \frac{\mathrm{d} q_{j}}{\mathrm{~d} t} \equiv \frac{\mathrm{d}(\delta q_{j})}{\mathrm{d} t}
\end{equation}

The virtual displacements and their time derivatives in the Lagrangian $ \delta L $ differentials variation replace by the total differential of $ d L $. However, from the definition there is no time increment in virtual displacements. Through the process of integration by parts, the time derivative of $ \delta q_j $ transfer to $ \partial L/\partial(dq_j/dt) $. This will allows Lagrangian derivatives to exchange $ d(\delta q_j)/dt $ to $ \delta q_j $ by independent virtual displacements factorization. The definite time integral $\delta I(L)$ set to zero and as $ \delta q_j $ are independent of the integrand, if and only equals zero by each coefficient of $ \delta q_j $ must also be zero. Furthermore, action functional integral increases with any variation of the functional. The action is stationary when its variation with respect to  $\Psi^{*}$ or $\Psi$ or with both is zero. As a consequence of this variation constrain, the integral $ I(L) $ change by the amount $\delta I(L)$ as,

\begin{equation}\label{eq-22}
\delta I(L)=\int_{t_{1}}^{t_{2}} d t \int_{V} \mathscr{L}_{\theta}(t, \Psi_{i}+\delta\Psi_{i}, \nabla \Psi_{i}+\delta \nabla \Psi_{i}, \dot\Psi_{i}+\delta \dot\Psi_{i}) d \boldsymbol{r} -\int_{t_{1}}^{t_{2}} d t \int_{V} \mathscr{L}_{\theta}(t, \Psi_{i}, \nabla \Psi_{i}, \dot\Psi_{i}) d \boldsymbol{r}
\end{equation}

The $\delta I(L)$ integral up to the first order $\delta \Psi_{i}$ parameter in,

\begin{equation}\label{eq-23}
\delta I(L)=\int_{t_{1}}^{t_{2}} d t  \int_{V} \sum_{i=1}^{n}\bigg\{\frac{\partial \mathscr{L}_{\theta}}{\partial \Psi_{i}} \delta \Psi_{i}+\sum_{j=1}^{3} \frac{\partial \mathscr{L}_{\theta}}{\partial(\partial \Psi_{i} / \partial \boldsymbol{r}_j)} \delta\Big(\frac{\partial \Psi_{i}}{\partial \boldsymbol{r}_j}\Big) +\frac{\partial \mathscr{L}_{\theta}}{\partial \dot\Psi_{i}} \delta \dot\Psi_{i} \bigg\} d \boldsymbol{r}
\end{equation}

From the generalized form of Green's theorem, 

\begin{equation}\label{eq-24}
\int_{V} \sum_{j=1}^{3} \frac{\partial \mathscr{L}_{\theta}}{\partial(\partial \Psi_{i} / \partial \boldsymbol{r}_j)} \frac{\partial \delta \Psi_{i} }{\partial \boldsymbol{r}_j} d\boldsymbol{r} = \int_{S} \delta \Psi_{i} \sum_{j=1}^{3} \frac{\partial \mathscr{L}_{\theta}}{\partial(\partial \Psi_{i} / \partial \boldsymbol{r}_j)} l_{j} d S -\int_{V} \delta \Psi_{i} \sum_{j=1}^{3} \frac{\partial}{\partial \boldsymbol{r}_j} \frac{\partial \mathscr{L}_{\theta}}{\partial(\partial \Psi_{i}/ \partial \boldsymbol{r}_j)} d \boldsymbol{r}
\end{equation}

Where the $l_{j}(j=1, \ldots, 3)$ are at any point cosines direction normal  to hyper-surface $S$ boundaries and further integration by parts yield,

\begin{equation}\label{eq-25}
\int_{t_{1}}^{t_{2}} \frac{\partial \mathscr{L}_{\theta}}{\partial \dot\Psi_{i}} \frac{\partial \delta \Psi_{i}}{\partial t} d t=\bigg[\delta \Psi_{i} \frac{\partial \mathscr{L}_{\theta}}{\partial \dot\Psi_{i}}\bigg]_{t_{1}}^{t_{2}}-\int_{t_{1}}^{t_{2}} \delta \Psi_{i} \frac{\partial}{\partial t}\Big(\frac{\partial \mathscr{L}_{\theta}}{\partial \dot\Psi_{i}}\Big) d t
\end{equation}

Since the $\delta \Psi_{i}(\boldsymbol{r}, t)$ vanish over the surface $S$ and at times $t=t_{1}, t_{2}$, Hence we obtain,

\begin{equation}\label{eq-26}
\delta I(L)=\int_{t_{1}}^{t_{2}} d t \int_{V} \sum_{i=1}^{n} \delta \Psi_{i}\bigg\{\frac{\partial \mathscr{L}_{\theta}}{\partial \Psi_{i}}-\sum_{j=1}^{3} \frac{\partial}{\partial \boldsymbol{r}_j} \frac{\partial \mathscr{L}_{\theta}}{\partial(\partial \Psi_{i} / \partial \boldsymbol{r}_j)}-\frac{\partial}{\partial t}\Big(\frac{\partial \mathscr{L}_{\theta}}{\partial \dot\Psi_{i}}\Big)\bigg\} d \boldsymbol{r}
\end{equation}

In terms of functional derivatives $\delta L$ of Lagrangian function $L$, from the integral of \cref{eq-26} is,

\begin{equation}\label{eq-27}
\delta L=\int_{V} \sum_{i=1}^{n}\bigg[\frac{\delta L}{\delta \Psi_{i}} \delta \Psi_{i}+\frac{\delta L}{\delta \dot\Psi_{i}} \delta \dot\Psi_{i}\bigg] d \boldsymbol{r}
\end{equation}

Where functional derivatives $\delta L$ with respect to $\Psi_{i}$ and $\dot\Psi_{i}$ are,

\begin{equation}\label{eq-28}
\frac{\delta L}{\delta \Psi_{i}}=\frac{\partial \mathscr{L}_{\theta}}{\partial \Psi_{i}}-\sum_{j=1}^{3} \frac{\partial}{\partial \boldsymbol{r}_j} \frac{\partial \mathscr{L}_{\theta}}{\partial(\partial \Psi_{i} / \partial \boldsymbol{r}_j)}
\end{equation}

\begin{equation}\label{eq-29}
\frac{\delta L}{\delta \dot\Psi_{i}}=\frac{\partial \mathscr{L}_{\theta}}{\partial \dot\Psi_{i}}-\sum_{j=1}^{3} \frac{\partial}{\partial \boldsymbol{r}_j} \frac{\partial \mathscr{L}_{\theta}}{\partial(\partial \dot\Psi_{i} / \partial \boldsymbol{r}_j)}
\end{equation}

The terms involving the $\partial \mathscr{L}_{\theta} / \partial(\partial \dot\Psi_{i} / \partial \boldsymbol{r}_j)$ on the right hand side of \cref{eq-29} being zero because $\mathscr{L}_{\theta}$ does not depend upon the spatial derivatives of $\dot\Psi_{i}$. Now from, \cref{eq-26}, if $I$ is stationary for arbitrary independent variations $\delta \Psi_{i}$, it follows from   $\delta I(L)=0$ as,

\begin{equation}\label{eq-30}
\begin{split}
\frac{\partial \mathscr{L}_{\theta}}{\partial \Psi_{i}}-\sum_{j=1}^{3} \frac{\partial}{\partial \boldsymbol{r}_j} \frac{\partial \mathscr{L}_{\theta}}{\partial(\partial \Psi_{i} / \partial \boldsymbol{r}_j)}-\frac{\partial}{\partial t}\Big(\frac{\partial \mathscr{L}_{\theta}}{\partial \dot\Psi_{i}}\Big)=0 \\
\frac{\partial \mathscr{L}_{\theta}}{\partial \Psi_{i}^{*}}-\sum_{j=1}^{3} \frac{\partial}{\partial \boldsymbol{r}_j} \frac{\partial \mathscr{L}_{\theta}}{\partial(\partial \Psi_{i}^{*} / \partial \boldsymbol{r}_j)}-\frac{\partial}{\partial t}\Big(\frac{\partial \mathscr{L}_{\theta}}{\partial \dot\Psi^{*}_{i}}\Big)=0
\end{split}
\end{equation}

Which are the Euler-Lagrange equations for $ \Psi_{i} $ and its complex conjugate $ \Psi^{*}_{i} $. For the Lagrangian density integral $ I(L) $, stationary state solution $\delta I(L)=0$ and from \cref{eq-28} and \cref{eq-29}, The stochastic differential equation (SDE) is,

\begin{equation}\label{eq-32}
\frac{\partial \mathscr{L}_{\theta}}{\partial \Psi_{i}}=\sum_{j=1}^{3} \frac{\partial}{\partial \boldsymbol{r}_j} \frac{\partial \mathscr{L}_{\theta}}{\partial(\partial \Psi_{i} / \partial \boldsymbol{r}_j)} \quad(i=1, \ldots, n)
\end{equation}

\cref{eq-32} is the Euler-Lagrange equations corresponding to the Lagrangian density, and by virtue of \cref{eq-17} provides a dynamical equation of the state. In the functional derivatives $\delta L$ form the \cref{eq-32} is,

\begin{equation}\label{eq-33}
\frac{\partial}{\partial t} \frac{\delta L}{\delta \dot\Psi_{i}}=\frac{\delta L}{\delta \Psi_{i}}
\end{equation}

This is analogous to Lagrange's equation if we regard the field component $\Psi_{i}$ at a particular point of space as one of the generalized coordinates $q_{r}$ of a dynamical system having an infinite number of degrees of freedom.

\subsection*{\texorpdfstring{Hamiltonian Density}{Hamiltonian Density}}

In the Hamiltonian formalism, canonical conjugate momentum $\Pi_{i}(\boldsymbol{r}, t)$ from the generalized coordinate $\Psi_{i}(\boldsymbol{r}, t)$ in the $3n$-dimensional configuration space from the Lagrangian density is,

\begin{equation}\label{eq-36}
\Pi_{i}(\boldsymbol{r}, t)\equiv \frac{\partial \mathscr{L}_{\theta}}{\partial \dot\Psi_{i}(\boldsymbol{r}, t)}=i \hbar \Psi^{*}_{i}(\boldsymbol{r}, t)
\end{equation}

Conjugate variables momentum $\Pi_{i}(\boldsymbol{r}, t)$ and generalized coordinate $\Psi_{i}(\boldsymbol{r}, t)$ pairs of operator that do not commute and incompatible observables. The operators $\widehat{\Psi}_{i}(\boldsymbol{r}, t)$ and $\widehat{\Pi}_{i}(\boldsymbol{r}, t)$, satisfy the canonical commutation relation as,

\begin{equation}\label{eq-36-1}
[\widehat{\Psi}_{i}(\boldsymbol{r}, t), \widehat{\Pi}_{i}(\boldsymbol{r}, t)]=\widehat{\Psi}_{i}(\boldsymbol{r}, t) \widehat{\Pi}_{i}(\boldsymbol{r}, t)-\widehat{\Pi}_{i}(\boldsymbol{r}, t) \widehat{\Psi}_{i}(\boldsymbol{r}, t)=i \hbar
\end{equation}

Furthermore, the generalized uncertainty principle between observables standard deviation $ \sigma_{\Psi_{i}(\boldsymbol{r}, t)} $ and $ \sigma_{\Pi_{i}(\boldsymbol{r}, t)}  $ and corresponding operators by definition as,

\begin{equation}\label{eq-36-2}
\sigma_{\Psi_{i}(\boldsymbol{r}, t)}^{2} \sigma_{\Pi_{i}(\boldsymbol{r}, t)}^{2} \geq\big(\frac{1}{2 i}\langle[\widehat{\Psi}_{i}(\boldsymbol{r}, t), \widehat{\Pi}_{i}(\boldsymbol{r}, t)]\rangle\big)
\end{equation}

Lagrangian density is defined upon generalized coordinate $\Psi_{i}(\boldsymbol{r}, t)$ and momentum $\Pi_{i}(\boldsymbol{r}, t)$; hence simultaneously, Legendre transformations of generalized coordinates and momentum do not influence both the form of Euler–Lagrange's equations and, consequently, Hamilton's equations. By taking the Legendre transformation of the Lagrangian density, $\mathscr{L}_{\theta}$, the Hamiltonian density $\mathcal{H}$ as a tensor function of generalized coordinates $\Psi_{i}(t)$ and conjugate canonical momenta $\Pi_{i}(t)$ is,

\begin{equation}\label{eq-35}
\mathcal{H}(\Psi(t),\Pi(t))=\sum_{i=1}^{n} \Pi_{i}(\boldsymbol{r}, t) \dot\Psi_{i}(\boldsymbol{r}, t)-\mathscr{L}_{\theta}(\Psi_{i}(t), \dot\Psi_{i}(t))=\sum_{i=1}^{n}\Psi^{*}_{i} \widehat{H} \Psi_{i}
\end{equation}

From the Hamiltonian tensor density $ \mathcal{H}$ of \cref{eq-35}, the Hamiltonian energy operator $\widehat{H}$ defines as the sum of the operators corresponding to the kinetic and potential energies of the state. For charge $q_{i}$ and mass $m_{i}$, for $(i=1,2,\ldots n)$ of $n$-particles system in the external electric field of $A_{0}(\boldsymbol{r}_{i},t)$ scalar potential, at the generalized coordinate $\boldsymbol{r}_{i}$, with canonical momenta operator $\widehat{\boldsymbol{P}}_{i}=-i\hbar\nabla_{i}=-i\hbar\partial/\partial \boldsymbol{r}_{i}$ conjugate with coordinate $\boldsymbol{r}_{i}$ fulfilling the canonical commutation relations. The gradient of $ i^{th} $ particle $ \nabla_{i} $, and potential energy is $V$.

The Hamiltonian operator $\widehat{H}$ is,

\begin{equation}\label{eq-34-1}
\widehat{H}=\sum_{i=1}^{n}\left\{\frac{-\hbar ^{2}}{2 m_{i}}[\nabla_{i}]^{2}+q_{i} A_{0}(\boldsymbol{r}_{i}, t)\right\}+V(\boldsymbol{r}_{1},\boldsymbol{r}_{2}, \ldots, \boldsymbol{r}_{n})
\end{equation}

Where $ \nabla_{i}^{2} $ is the Laplacian operator in $ i^{th} $ particle coordinates, $(\boldsymbol{r}_{1},\boldsymbol{r}_{2},\ldots \boldsymbol{r}_{n})$ is $3n$-dimensional vector in configuration space. In the $n$-identical particles Fermionic system, potential  $V$ is the sum of one $V^{(1)}(\boldsymbol{r}_{i})$ and a symmetric two-body $V^{(2)}(\boldsymbol{r}_{i}, \boldsymbol{r}_{j})$ potentials energy as,

\begin{equation}\label{eq-34-3}
V(\boldsymbol{r}_{1},\boldsymbol{r}_{2},\ldots, \boldsymbol{r}_{n})=\sum_{i=1}^{n}V^{(1)}(\boldsymbol{r}_{i})+\frac{1}{2}\sum_{i\neq j=1}^{n}V^{(2)}(\boldsymbol{r}_{i},\boldsymbol{r}_{j})
\end{equation}

Where $V^{(2)}(\boldsymbol{r}_{i}, \boldsymbol{r}_{j})=V^{(2)}(\boldsymbol{r}_{j}, \boldsymbol{r}_{i})$. 
Now, spatial integration of Hamiltonian energy density in the $3n$-dimensional volume is,

\begin{equation}\label{eq-34}
H=\int_{V} \mathcal{H} d \boldsymbol{r} = \sum_{i=1}^{n} \int \Psi^{*}_{i}(\boldsymbol{r}, t) \widehat{H} \Psi_{i}(\boldsymbol{r}, t) d^{3n}\boldsymbol{r}
\end{equation}

Where the integration over the volume of entire configuration space with element of $d^{3n}\boldsymbol{r}= d^{3}\boldsymbol{r}_{1} d^{3}\boldsymbol{r}_{2}\ldots d^{3}\boldsymbol{r}_{n}$. 
By using canonical momenta density from \cref{eq-36} in the Hamiltonian density of the system \cref{eq-34}, The Hamiltonian functional $\mathbb{H}[\Psi_{i}(t),\Pi_{i}(t)]$ defines as, 

\begin{equation}\label{eq-34-2}
\mathbb{H}[\Psi_{i}(t), \Pi_{i}(t)]=\int_{V}  \mathcal{H}(t) d \boldsymbol{r}=\sum_{i=1}^{n} \frac{1}{i \hbar} \int \Pi_{i}(\boldsymbol{r}, t) \widehat{H} \Psi_{i}(\boldsymbol{r}, t) d^{3n} \boldsymbol{r}
\end{equation}

The variation in $H$ arising from independent increments in the field components $\Psi_{i}$ and their time derivatives $\dot\Psi_{i}$, alternatively, in the $\Psi_{i}$ and the momentum densities $\Pi_{i}$ is given as,

\begin{equation}\label{eq-37}
\delta H=\int_{V} \sum_{i=1}^{n}\bigg\{\Pi_{i} \delta \dot\Psi_{i}+\dot\Psi_{i} \delta \Pi_{i}-\frac{\delta L}{\delta \Psi_{i}} \delta \Psi_{i}-\frac{\delta L}{\delta \dot\Psi_{i}} \delta \dot\Psi_{i}\bigg\} d \boldsymbol{r}
\end{equation}

Using equations \cref{eq-33} and \cref{eq-36} in the abstract notation form as,

\begin{equation}\label{eq-38}
\delta H=\int_{V} \sum_{i=1}^{n}\big[\dot\Psi_{i} \delta \Pi_{i}-\dot{\Pi}_{i} \delta \Psi_{i}\big] d \boldsymbol{r}
\end{equation}

The Hamiltonian density $ \mathcal{H}$ as a function of the field components $\Psi_{i}$, canonical momentum densities $\Pi_{i}$, spacial derivatives $\partial \Psi_{i} / \partial \boldsymbol{r}_j$, and time $t$, the variation $ \mathcal{H}$ is,

\begin{equation}\label{eq-39}
\delta H=\int_{V} \sum_{i=1}^{n}\bigg\{\frac{\partial \mathcal{H}}{\partial \Psi_{i}} \delta \Psi_{i}+\frac{\partial \mathcal{H}}{\partial \Pi_{i}} \delta \Pi_{i}+\sum_{j=1}^{3} \frac{\partial \mathcal{H}}{\partial(\partial \Psi_{i} / \partial \boldsymbol{r}_j)} \delta\Big(\frac{\partial \Psi_{i}}{\partial \boldsymbol{r}_j}\Big)\bigg\} d \boldsymbol{r}
\end{equation}

\begin{equation}\label{eq-40}
=\int_{V} \sum_{i=1}^{n}\bigg\{\Big(\frac{\partial \mathcal{H}}{\partial \Psi_{i}}-\sum_{j=1}^{3} \frac{\partial}{\partial \boldsymbol{r}_j} \frac{\partial \mathcal{H}}{\partial(\partial \Psi_{i} / \partial \boldsymbol{r}_j)}\Big) \delta \Psi_{i}+\frac{\partial \mathcal{H}}{\partial \Pi_{i}} \delta \Pi_{i}\bigg\} d \boldsymbol{r}
\end{equation}

Using Green's theorem and the boundary conditions imposed on $\delta \Psi_{i}$ are similar to those introduced in the Lagrangian density discussed in the previous section. 

In the abstract notation form expressed as,

\begin{equation}\label{eq-41}
\delta H=\int_{V} \sum_{i=1}^{n}\Big[\frac{\delta H}{\delta \Psi_{i}} \delta \Psi_{i}+\frac{\delta H}{\delta \Pi_{i}} \delta \Pi_{i}\Big] d \boldsymbol{r}
\end{equation}

Where,

\begin{equation}\label{eq-42}
\frac{\delta H}{\delta \Psi_{i}}=\frac{\partial \mathcal{H}}{\partial \Psi_{i}}-\sum_{j=1}^{3} \frac{\partial}{\partial \boldsymbol{r}_j} \frac{\partial \mathcal{H}}{\partial(\partial \Psi_{i} / \partial \boldsymbol{r}_j)}
\end{equation}

And, 

\begin{equation}\label{eq-43}
\frac{\delta H}{\delta \Pi_{i}}=\frac{\partial \mathcal{H}}{\partial \Pi_{i}}-\sum_{j=1}^{3} \frac{\partial}{\partial \boldsymbol{r}_j} \frac{\partial \mathcal{H}}{\partial(\partial \Pi_{i} / \partial \boldsymbol{r}_j)}
\end{equation}

are the functional derivatives of $H$ with respect to $\Psi_{i}$ and $\Pi_{i}$. As $\mathcal{H}$ does not depend upon the spatial derivatives of $\Pi_{i}$, hereabouts the second term on the right-hand side of \cref{eq-43} vanishes. 
By comparing \cref{eq-38} with \cref{eq-41}, we now see that,

\begin{equation}\label{eq-44}
\dot{\Psi}_{i}=\frac{\delta H}{\delta \Pi_{i}}
\end{equation}
\begin{equation}\label{eq-45}	
\dot{\Pi}_{i}=-\frac{\delta H}{\delta \Psi_{i}}
\end{equation}

Which closely resemble Hamilton's equations of classical dynamics. We conclude here by examining the condition for the Hamiltonian function $H$ to conserve energy. By taking the differential of $H$ with respect to time, and following from the field equations \cref{eq-44} and \cref{eq-45}, we obtain,

\begin{equation}\label{eq-46}
\frac{d H}{d t}=\int_{V}\bigg\{\sum_{i=1}^{n}\Big[\frac{\delta H}{\delta \Psi_{i}} \dot{\Psi}_{i}+\frac{\delta H}{\delta \Pi_{i}} \dot{\Pi}_{i}\Big]+\frac{\partial \mathcal{H}}{\partial t}\bigg\} d \boldsymbol{r}=\int_{V} \frac{\partial \mathcal{H}}{\partial t} d \boldsymbol{r}
\end{equation}

Here, if the Hamiltonian density $\mathcal{H}$ does not depend explicitly on time $t$, then $H$ is constant, and for that we must have condition,

\begin{equation}\label{eq-48}
\frac{d H}{d t}=0
\end{equation}

For the completeness we will consider the Hamiltonian density $\mathcal{H}$ and introducing the momenta $ {\Pi}_{i} $ and $ {\Pi}^{*}_{i} $ canonically conjugate to $\Psi_{i}$ and ${\Psi}^{*}_{i}$ respectively as,

\begin{equation}\label{eq-53}
{\Pi}_{i}=\frac{\partial \mathscr{L}_{\theta}}{\partial \dot{\Psi}_{i}}, \quad {\Pi}^{*}_{i}=\frac{\partial \mathscr{L}_{\theta}}{\partial \dot{\Psi}^{*}_{i}}
\end{equation}

And, the momenta $\Pi_{i}^{(1)} $ and $ \Pi_{i}^{(2)}$ canonically conjugate to $\Psi_{i}^{(1)}$ and $\Psi_{i}^{(2)}$ respectively as,

\begin{equation}\label{eq-54}
\Pi_{i}^{(1)}=\frac{\partial \mathscr{L}_{\theta}}{\partial \dot{\Psi}_{i}^{(1)}}, \quad \Pi_{i}^{(2)}=\frac{\partial \mathscr{L}_{\theta}}{\partial \dot{\Psi}_{i}^{(2)}}
\end{equation}

So that,
\begin{equation}\label{eq-55}
\Pi_{i}^{(1)}=\frac{1}{\sqrt{2}}\big[\Pi_{i}+{\Pi}^{*}_{i}\big], \;\;
\Pi_{i}^{(2)}=\frac{i}{\sqrt{2}}\big[\Pi_{i}-{\Pi}^{*}_{i}\big]
\end{equation}

Therefore, the Hamiltonian density $\mathcal{H}$ is,

\begin{equation}\label{eq-56}
\mathcal{H}=\sum_{i=1}^{n}\big[{\Pi}_{i}^{(1)} \dot{\Psi}_{i}^{(1)}+\Pi_{i}^{(2)} \dot{\Psi}_{i}^{(2)}\big]-\mathscr{L}_{\theta}=\sum_{i=1}^{n}\big[\Pi_{i} \dot{\Psi}_{i}+{\Pi}^{*}_{i} \dot{\Psi}^{*}_{i}\big]-\mathscr{L}_{\theta}
\end{equation}

Finally, we should note that unless the Lagrangian density $\mathscr{L}_{\theta}$ is real, ${\Pi}^{*}_{i}$ will not be the complex conjugate of $\Pi_{i}$.

\subsection*{\texorpdfstring{Electronic Charge Density}{Electronic Charge Density}}

The electronic charge density $\rho$ associated with the Lagrangian density of complex particle field defines as,

\begin{equation}\label{eq-49}
\rho=-i \epsilon \sum_{i=1}^{n}\Big[\frac{\partial \mathscr{L}_{\theta}}{\partial \dot{\Psi}_{i}} \Psi_{i}-\frac{\partial \mathscr{L}_{\theta}}{\partial \dot{\Psi}^{*}_{i}} {\Psi}^{*}_{i}\Big]
\end{equation}

\subsection*{\texorpdfstring{Electronic Current Flux }{Electronic Current Flux}}

The associated electronic current vector density $\boldsymbol{j}$ components are defined as,

\begin{equation}\label{eq-50}
\boldsymbol{j}_{j}=-i \epsilon \sum_{i=1}^{n}\bigg[\frac{\partial \mathscr{L}_{\theta}}{\partial(\partial \Psi_{i} / \partial \boldsymbol{r}_j)} \Psi_{i}-\frac{\partial \mathscr{L}_{\theta}}{\partial(\partial {\Psi}^{*}_{i} / \partial \boldsymbol{r}_j)} {\Psi}^{*}_{i}\bigg] \quad(j=1,2,3)
\end{equation}

Where $\epsilon$ is a real number.

\subsection*{\texorpdfstring{Current Continuity Equation}{Current Continuity Equation}}

Noether's theorem states that the conservation law of nature holds in the global configuration space of a physically symmetric environment through infinitesimal gauge transformations operation.\cite{noether_invariante_1918, Schrodinger_Quantisierung4_1926} In configuration space, invariance in Lagrangian density by using the field equation \cref{eq-30} presents an equation of continuity for the probability species as,

\begin{equation}\label{eq-51}
\begin{aligned}
\frac{\partial \rho}{\partial t}+\operatorname{div} \boldsymbol{j}=-i \epsilon \sum_{i=1}^{n}\bigg[\Big\{\frac{\partial \mathscr{L}_{\theta}}{\partial \Psi_{i}} \Psi_{i}+\sum_{j=1}^{3} \frac{\partial \mathscr{L}_{\theta}}{\partial(\partial \Psi_{i} / \partial \boldsymbol{r}_j)} \frac{\partial \Psi_{i}}{\partial \boldsymbol{r}_j}+\frac{\partial \mathscr{L}_{\theta}}{\partial \dot{\Psi}_{i}} \dot{\Psi}_{i}\Big\} -\Big\{\frac{\partial \mathscr{L}_{\theta}}{\partial {\Psi}^{*}_{i}} {\Psi}^{*}_{i}+\sum_{j=1}^{3} \frac{\partial \mathscr{L}_{\theta}}{\partial(\partial {\Psi}^{*}_{i} / \partial \boldsymbol{r}_j)} \frac{\partial {\Psi}^{*}_{i}}{\partial \boldsymbol{r}_j}+\frac{\partial \mathscr{L}_{\theta}}{\partial \dot{\Psi}^{*}_{i}} \dot{\Psi}^{*}_{i}\Big\}\bigg]
\end{aligned}
\end{equation}

Here we assume that the Lagrangian density from \cref{eq-17} is invariant under the gauge transformation. From this particular invariant case of Noether theorem and from the calculus of variations,

\begin{equation}\label{eq-52}
\frac{\partial \rho}{\partial t}+\operatorname{div} \boldsymbol{j}=0
\end{equation}

Therefore, the continuity equation satisfies separately both $\rho$ and $\boldsymbol{j}$.

\subsection*{\texorpdfstring{Schr{\"o}dinger Equation}{Schr{\"o}dinger Equation}}

The Lagrangian density $\mathscr{L}_{\theta}$ in terms of time-dependent trial wave-function $\Psi(t)$, and its complex conjugate ${\Psi}^{*}(t)$ for the non-relativistic schr{\"o}dinger wave is,

\begin{equation}\label{eq-58-0}
\mathscr{L}_{\theta}=\Psi^{*}(t)\big[i \hbar \partial_{t}-\widehat{H}\big] \Psi(t)
\end{equation}

Where $\widehat{H}$ is aforementioned Hamiltonian operator from \cref{eq-34-1}. 

Using Hamilton's principle of Least or stationary Action on the Lagrangian density, the system's dynamical solution derives using the variational principle method. The Action functional $\delta \mathcal {S}[\Psi^{*},\Psi]$ for the Lagrangian density $\mathscr{L}_{\theta}$ is stationary for,

\begin{equation}\label{eq-58-1}
\delta \mathcal {S}[\Psi^{*}, \Psi] \stackrel{\text{def}}{=} \delta I(L)= \int_{t_{1}}^{t_{2}} \delta L \;\; dt = \int_{T} dt \int \delta\mathscr{L}_{\theta}(\Psi,\Psi^{*},\ldots)d^{3n} \boldsymbol{r}=0\rightarrow\text{stationary}
\end{equation}

Where the integration is over all time and entire configuration space with an element of volume $d^{3n}\boldsymbol{r}=d^{3}\boldsymbol{r}_{1}\ldots d^{3}\boldsymbol{r}_{n}$. 
The action is stationary when $\Psi^{*}$ or $\Psi$ or both variation is zero and the boundary conditions at infinity is when $ r_{i}\rightarrow\infty$ and $|t|\rightarrow\infty$, the variation $\delta\Psi$ and $\delta\Psi^{*}$ again vanishes. Putting these condition and from \cref{eq-58-1} variational principle, The Lagrangian density $\mathscr{L}_{\theta}$ from \cref{eq-58-0} of schr{\"o}dinger equation is,

\begin{equation}\label{eq-58}
\mathscr{L}_{\theta}=-\frac{\hbar^{2}}{2 m} \nabla {\Psi}^{*}\cdot \nabla \Psi-V(\boldsymbol{r}, t) {\Psi}^{*} \Psi+\frac{1}{2} i \hbar\big[{\Psi}^{*} \dot{\Psi}-\Psi \dot{\Psi}^{*}\big]
\end{equation}

Where a particle of mass $m$ moving under the Action of a potential $V(\boldsymbol{r}, t)$. Particle field $\Psi$ with its complex component ${\Psi}^{*}$ independent variations yield the conjugate Schr{\"o}dinger wave \cref{eq-57} as,

\begin{equation}\label{eq-57}
\begin{aligned}
-\frac{\hbar^{2}}{2 m} \nabla^{2} \Psi+V(\boldsymbol{r}, t) \Psi&=i \hbar \frac{\partial \Psi}{\partial t} \\
-\frac{\hbar^{2}}{2 m} \nabla^{2} {\Psi}^{*}+V(\boldsymbol{r}, t) {\Psi}^{*}&=-i \hbar \frac{\partial {\Psi}^{*}}{\partial t}
\end{aligned}
\end{equation}

Introducing the canonically conjugate momenta $ \Pi $ and $ \Pi^{*} $ to $\Psi$ and ${\Psi}^{*}$ as,

\begin{equation}\label{eq-60}
\begin{aligned}
\Pi&=\frac{\partial \mathscr{L}_{\theta}}{\partial \dot{\Psi}}=\frac{1}{2} i \hbar {\Psi}^{*} \\
{\Pi}^{*}&=\frac{\partial \mathscr{L}_{\theta}}{\partial \dot{\Psi}^{*}}=-\frac{1}{2} i \hbar \Psi
\end{aligned}
\end{equation}

The Hamiltonian density $ \mathcal{H} $ is,

\begin{equation}\label{eq-61}
\mathcal{H} =\Pi \dot{\Psi}+{\Pi}^{*} {\dot\Psi}^{*}-\mathscr{L}_{\theta} =\frac{\hbar^{2}}{2 m} \nabla {\Psi}^{*} \cdot \nabla \Psi+V(\boldsymbol{r}, t) {\Psi}^{*} \Psi
\end{equation}

The Hamiltonian density $\mathcal{H}$ in the abstract form, 

\begin{equation}\label{eq-62}
\mathcal{H}=\frac{1}{i \hbar}\bigg\{\frac{\hbar^{2}}{2 m}\big[\nabla \Pi \cdot \nabla \Psi-\nabla {\Pi}^{*} \cdot \nabla {\Psi}^{*}\big]+V(\boldsymbol{r}, t)\big[\Pi \Psi-{\Pi}^{*} {\Psi}^{*}\big]\bigg\}
\end{equation}

Where \cref{eq-62}, is conjugate form of Schr{\"o}dinger equation, and functional derivatives defines as $\delta H / \delta \Pi$ and $\delta H / \delta {\Pi}^{*}$, as Hamiltonian density $\mathcal{H}$ depends upon the gradients of the canonical momenta, by using Hamilton's equations as, 

\begin{equation}\label{eq-63}
\dot{\Psi}=\frac{\delta H}{\delta \Pi}, \quad \dot{\Pi}=-\frac{\delta H}{\delta \Psi} ; \quad \dot{\Psi}^{*}=\frac{\delta H}{\delta {\Pi}^{*}}, \quad \dot{\Pi}^{*}=-\frac{\delta H}{\delta {\Psi}^{*}}
\end{equation}

Again, Noether's theorem states that the conservation law of nature holds in the global configuration space of a physically symmetric environment through infinitesimal gauge transformations operation. \cite{noether_invariante_1918, Schrodinger_Quantisierung4_1926} In configuration space, invariance in Lagrangian density by using the field equation \cref{eq-30} presents an equation of continuity for the probability species.

The variation in Lagrangian density $\delta\mathscr{L}_{\theta}$ from \cref{eq-58-0} and \cref{eq-58} with respect to $\delta\Psi$ is, 

\begin{equation}\label{eq-58-4}
\delta\mathscr{L}_{\theta}=i \hbar\frac{\partial}{\partial t} \big[{\Psi}^{*} \delta \Psi\big]+\frac{i \hbar}{2m} \nabla \cdot\big[{\Psi}^{*} \nabla \delta\Psi-\delta\Psi \nabla {\Psi}^{*}\big]
\end{equation}

Global gauge transformation changes the wave-function $\Psi$ by an infinitesimal constant. Furthermore, the externally applied scalar electric potential remained unchanged due to a constant gauge factor and canceled out in gauge transformation. Therefore by equating the variation in Lagrangian density $\delta\mathscr{L}_{\theta}$ equal to zero, a generalized continuity equation for probability wave-function $\Psi,\Psi^{*}$ is,

\begin{equation}\label{eq-65-1}
\delta \mathcal {S}[\Psi^{*}, \Psi] = \delta I(L)= \int_{t_{1}}^{t_{2}} \delta L \;\; dt = \int_{T} dt \int \delta\mathscr{L}_{\theta}(\Psi,\Psi^{*},\ldots)d^{3n} \boldsymbol{r}=0\rightarrow\text{stationary}
\end{equation}

\begin{equation}\label{eq-65}
\begin{aligned}
\delta\mathscr{L}_{\theta}= \frac{\partial}{\partial t} \int {\Psi}^{*} \Psi d \boldsymbol{r} - \frac{i \hbar}{2 m} \int \nabla \cdot\big[{\Psi}^{*} \nabla \Psi-\Psi \nabla {\Psi}^{*}\big] d \boldsymbol{r} = 0\\ \int\Big[{\Psi}^{*} \frac{\partial \Psi}{\partial t}+\Psi \frac{\partial {\Psi}^{*}}{\partial t}\Big] d \boldsymbol{r} = \frac{i \hbar}{2 m} \int\Big[{\Psi}^{*} \nabla^{2} \Psi-\Psi \nabla^{2} {\Psi}^{*}\Big] d \boldsymbol{r} 
\end{aligned}
\end{equation}

For single Fermionic wave-function $\phi(\boldsymbol{r}, t)$ state, the Action functional is, 

\begin{equation}\label{eq-65-2}
\begin{aligned}
\mathcal {S}[\phi^{*},\phi]\stackrel{\text{def}}{=}N \int dt\int d^{3}\boldsymbol{r}\phi^{*}(\boldsymbol{r}, t)&\bigg\{i\hbar\partial_{t}\phi(\boldsymbol{r}, t)-\frac{1}{2m}[-i\hbar\nabla]^{2}\phi(\boldsymbol{r}, t)-q{A}_{0}\phi(\boldsymbol{r}, t) -V^{(1)}(\boldsymbol{r})\phi(\boldsymbol{r}, t) \\
&-\frac{1}{2}(N-1)\int d^{3}\boldsymbol{r}'V^{(2)}(\boldsymbol{r}, \boldsymbol{r}')\phi^{*}(\boldsymbol{r}', t)\phi(\boldsymbol{r}', t)\phi(\boldsymbol{r}, t)\bigg\}=0 \rightarrow\text{stationary}
\end{aligned}
\end{equation}

Variation of Action functional  $ \delta \mathcal {S}[\phi^{*},\phi] $  with respect to $\delta\phi^{*}$, gives $n$-particles time-dependent Hartree approximation of a nonlinear Schr{\"o}dinger wave in the external electric field of a scalar potential $A_{0}(\boldsymbol{r},t)$ as, 

\begin{equation}\label{eq-65-3}
\begin{aligned}
i \hbar\frac{\partial}{\partial t}\phi(\boldsymbol{r}, t)=\frac{1}{2m}[-i\hbar\nabla]^{2}\phi(\boldsymbol{r}, t)+[qA_{0}(\boldsymbol{r},t)+V^{(1)}(\boldsymbol{r})]\phi(\boldsymbol{r}, t) +(N-1)\displaystyle \int d^{3}\boldsymbol{r}'V^{(2)}(\boldsymbol{r}, \boldsymbol{r}')\phi^{*}(\boldsymbol{r}', t)\phi(\boldsymbol{r}', t)\phi(\boldsymbol{r}, t)
\end{aligned}
\end{equation}

The nonlinear term arising is the average potential on one particle due to all the other particles in the system. In quantum mechanics, the wave-function $\Psi$ is interpreted as ${\Psi}^{*} \Psi d \boldsymbol{r}$ is equal to the probability of observing the particle at time $t$ within the volume element $d \boldsymbol{r}$ surrounding the point with position vector $\boldsymbol{r}$. Therefore, ${\Psi}^{*}\Psi$ is defined as the probability density.
From \cref{eq-49}, putting $\epsilon=e / \hbar$, the probability charge density is,

\begin{equation}\label{eq-64}
\rho=-\frac{i e}{\hbar}\Big[\frac{\partial \mathscr{L}_{\theta}}{\partial \dot{\Psi}} \Psi-\frac{\partial \mathscr{L}_{\theta}}{\partial \dot{\Psi}^{*}} {\Psi}^{*}\Big]=e {\Psi}^{*} \Psi
\end{equation}

Where $e$ is the charge of the particle. Following the above definition of probability charge density and from the variation in the Lagrangian density $\delta\mathscr{L}_{\theta}$ of \cref{eq-65}

\begin{equation}\label{eq-66}
\int\Big\{\frac{\partial \rho}{\partial t}+\operatorname{div} \boldsymbol{j}\Big\} d \boldsymbol{r}=0
\end{equation}

Where $ \boldsymbol{j} $ is probability electric current density vector define as,

\begin{equation}\label{eq-67}
\boldsymbol{j}=-\frac{i e \hbar}{2 m}\big[{\Psi}^{*} \nabla \Psi-\Psi \nabla {\Psi}^{*}\big]
\end{equation}

Where operator $ -i\hbar\nabla =-i\hbar\partial/\partial \boldsymbol{r} $ is canonical momenta operator conjugate to the coordinate $\boldsymbol{r}$ and gives the velocity flux for probability wave-function $\Psi,\Psi^{*}$. 
Therefore, the electric current density vector components from \cref{eq-50} in terms of the Lagrangian density from \cref{eq-58} is expressed after substituting $\epsilon=e / \hbar$ as,

\begin{equation}\label{eq-68}
\boldsymbol{j}_{j}=-\frac{i e}{\hbar}\bigg[\frac{\partial \mathscr{L}_{\theta}}{\partial(\partial \Psi / \partial \boldsymbol{r}_j)} \Psi-\frac{\partial \mathscr{L}_{\theta}}{\partial(\partial {\Psi}^{*} / \partial \boldsymbol{r}_j)} {\Psi}^{*}\bigg]
\end{equation}

The continuity equation relating the charge and current densities from \cref{eq-66} in the arbitrary volume of integration is, \cite{Schrodinger_Quantisierung4_1926}

\begin{equation}\label{eq-69}
\frac{\partial \rho}{\partial t}+\operatorname{div} \boldsymbol{j}=0
\end{equation}

To efficiently learn multi-layer perceptron approximate the Lagrangian density with a finite-difference Lagrangian operator on a grid spacing $\Delta x$ as, 

\begin{equation}\label{eq-13}
\mathscr{L}_{\theta_i} = \boldsymbol{\dot \phi_i}^2 - \bigg[\frac{\boldsymbol{\phi}_{i+1} - \boldsymbol{\phi}_{i-1}}{2 \Delta x}\bigg]^2
\end{equation}

\subsection*{\texorpdfstring{Partition Function, Expectation Value, and Entropy}{Partition Function, Expectation Value, and Entropy}}

The partition function defined the likelihood of specific configuration values from the constitution of the configuration of a system as a probability measure on the probability space. From the statistical mechanics, this measure is also called the Gibbs measure. In the quantum field theory generating functional is analogous to the partition function of statistical mechanics. If an exact closed-form expression is solved will completely define everything about the system, possibly we can know about a system. In the partition function, artificial auxiliary functions are introduced to obtain the random variables' expectation value $ { \mathbb {E} } $. These auxiliary functions refer to source fields in quantum field theory's path integral formulation. Multiple differentiation of these random variables leads to connected correlation functions. The quantum field theory deduces these connected correlation functions as Green's function for the differential operator as Fredholm kernels on Hilbert space and refers to them as propagators. Intuitively, the expected value of a random variable is a generalized weighted average arithmetic mean of a large number of the independent variable. In physics, we also called the expected value the first moment. If the configuration constitutes space, there is one configuration for which expectation value probability is maximum; from the physics point of view, we called it ground state. Furthermore, if this is a unique configuration, it is called a non-degenerate ground state, and the system is ergodic.

Von Neumann \textit{et al.} introduces the Gibbs classical entropy concepts into quantum mechanics to develop quantum statistical mechanics. The entropy quantifies the system's departure from a pure state to the degree of mixing of the state for a finite-dimensional matrix representation. A quantum-mechanical system of a density matrix  $ \rho $, the Von Neumann entropy $ \mathbb{S} $ is,

\begin{equation}\label{eq-70}
{ \mathbb{S}(\rho )=- {Tr} (\rho \ln \rho )}
\end{equation}

Where $ {Tr} $  denotes the trace of matrix and $ \ln $ is the natural matrix logarithm. Defining the density matrix  $ \rho $ in terms of its discrete eigenvectors $ { |1\rangle ,|2\rangle ,|3\rangle ,\dots } $ is,

\begin{equation}\label{eq-71}
{ \rho =\sum _{i}\eta _{i}|i\rangle \langle i|}
\end{equation}

From \cref{eq-71}, into \cref{eq-70}, the Von Neumann  entropy is, which is equivalent to Shannon information theoretic entropy,\cite{shannon1948a,shannon1948a2}

\begin{equation}\label{eq-72}
{ \mathbb{S}(\rho )=-\sum _{i}\eta _{i}\ln \eta _{i}}
\end{equation}

Von Neumann's joint entropy of a composite system can be lower than the entropy of any of its constituent parts if components are entangled, as in the case of Bell state as a pure state with zero entropy. However, Shannon's \textit{et al.} information-theoretic entropy always add-up and is higher than the components.

\subsection*{\texorpdfstring{Principle of Maximum Entropy}{Principle of Maximum Entropy}}

Maxwell \textit{et al.} presented the kinetic theory of gases in 1867 on-premises of molecular chaos approximation in physics based on his first work in 1860. \cite{maxwell_1860ii,maxwell_1860v,maxwell_1867iv} The underlying assumption is that colliding particles' velocities are uncorrelated and independent of their position. Therefore, considering each particle separately, the collision event probabilities can be evaluated from the given colliding pair particles' velocities. In a small region of $\delta r $ ignore finding any other correlation between the probability of one particle with velocity $v$ and another velocity $v'$. This assumption asserts a physical hypothesis that entering characterizing particles' distribution function can be factorized in a collision of particles. The principle of maximum entropy relates to molecular chaos or "stosszahlansatz" in the writings of Ehrenfest \textit{et al.} and heuristically interprets the most probable configuration of particles before the collision.\cite{ehrenfest1959the} Furthermore, from the information theory perspective, it can be inference as the knowledge about a system's current state is best represented by probability distribution with the largest entropy as the maximum entropy distribution makes the fewest assumptions about the true distribution of the system's states. In this regard, the maximum entropy principle can be viewed as an application of Occam's razor principle of theory construction.

\subsection*{\texorpdfstring{Principle of Maximum Caliber}{Principle of Maximum Caliber}}

In the context of non-equilibrium statistical mechanics, Jaynes \textit{et al.} \cite{jaynes1980the} generalized the principle of maximum entropy to postulate the maximum path entropy or maximum caliber based on the hypothesis that the most unbiased probability distribution of paths is the one that maximizes their Shannon entropy over the paths space. From \cref{eq-1}, and \cref{eq-18} the classical $ \mathscr{L}_{\theta}(\boldsymbol{q}(t),\boldsymbol{\dot q}(t),t) $ and quantum Lagrangian density  $ \mathscr{L}_{\theta}\Big(t, \Psi_{i}, \Psi_{i}^{*}, \nabla \Psi_{i}, \nabla \Psi_{i}^{*}, \dot\Psi_{i}, \dot\Psi_{i}^{*}\Big) $, the entropy of paths defines as the caliber $ \mathit{S}$ of the n-dynamical constraints $ (\boldsymbol{q}(t),\boldsymbol{\dot q}(t),t) $, or  $\Psi_{i}(\boldsymbol{r},t)$ system in the time interval $ { t\in [0,T]} $ define by the path integral formulation for the Lagrangian density integral $ I(L) $, in a non-equilibrium statistical systems representation with n-degrees of freedom, the caliber $ \mathit{S}$ is

\begin{equation}\label{eq-73}
{ \mathit{S}[I [L(\boldsymbol{q}(t),\boldsymbol{\dot q}(t),t)]]=\int D_{\boldsymbol{q}}I [L(\boldsymbol{q}(t),\boldsymbol{\dot q}(t),t)]\,\ln {I [L(\boldsymbol{q}(t),\boldsymbol{\dot q}(t),t)] \over \pi [\boldsymbol{q}(t),\boldsymbol{\dot q}(t),t]}}
\end{equation}

With, the probability density functional as, 

\begin{equation}\label{eq-74}
{ I [L(\boldsymbol{q}(t),\boldsymbol{\dot q}(t),t)]=\exp \Bigg\{-\sum _{i=0}^{n}\int _{0}^{T}dt\,\,\ell_{\theta_i}(t)\mathscr{L}_{\theta_i}(\boldsymbol{q}(t),\boldsymbol{\dot q}(t),t)\Bigg\}}
\end{equation}

Where for n-constraint Lagrangian density integral $ I(L) $ path integral expectation value is,

\begin{equation}\label{eq-75}
{ \int D_{\boldsymbol{q}}I [L(\boldsymbol{q}(t),\boldsymbol{\dot q}(t),t)]\mathscr{L}_{\theta_i}(\boldsymbol{q}(t),\boldsymbol{\dot q}(t),t)=\langle \mathscr{L}_{\theta_i}(\boldsymbol{q}(t),\boldsymbol{\dot q}(t),t) \rangle =\ell (t)}
\end{equation}

The relative change in probability distribution in the system's entropy over a time currently away from the maximum entropy thermodynamic equilibrium state defines through the fluctuation theorem of time-averaged irreversible entropy production in non-equilibrium statistical mechanics, With the increases in the arrow of time or system size, the probability of observing an entropy production opposite to that dictated by the second law of thermodynamics decreases exponentially. The Kullback-Leibler divergence uses to define relative entropy difference over time. On the same probability space, $ { {\mathcal {Q}}} $, for two discrete $ { P} $ and $ { Q} $ probability distributions, the relative entropy from $ {Q} $ to $ { P} $ in Kullback-Leibler divergence is,

\begin{equation}\label{eq-76}
{ D_{\text{KL}}(P\parallel Q)=\sum _{\boldsymbol{q}\in {\mathcal {Q}}}P(\boldsymbol{q})\log ({\frac {P(\boldsymbol{q})}{Q(\boldsymbol{q})}})}
\end{equation}

\subsection*{\texorpdfstring{Discussion on the Functional Neural Networks Topologies}{Discussion on the Functional Neural Networks Topologies}}

In this decade, with the efficient Deep learning neural network implementation through graphics processing unit, tensor processing unit, hardware realization, and high-level application programming interface libraries advancement, we envision the development of physics-inspired domain-specific functional neural network topologies, e.g.,  fluctuation-dissipation response neural network for response theory of perturbation in transport, \cite{kirkwood_statistical_1946,kirkwood_statistical_1947,kubo_statistical-mechanical_1957,kubo_statistical-mechanical_1957-1} relaxation time neural network for Boltzmann transport, and Feynman propagator neural network will emerge for the specific task of solving the dynamical problem as a co-processor in synergy with conventional equation-based mathematical solvers. We envision this thrust into the new interdisciplinary subject area, e.g., ``Quanta-Informatics'', ``Mecha-Informatics'', ``Physico-Informatics'', and ``Chemo-Informatics'' a few shortly after the adoption of artificial neural network-based informatics and stochastic inference techniques in the conventional natural science field. This data-enabled prediction paradigm shift will transform the way we do scientific calculations in the past and throw new insight into nature dynamics. In the past few years, the initial response of the electronic structure community in quantum physics for the DL/ML neural network technique was the design of different descriptors to feature engineering by encoding the local microscopic environment of the atomic structure in different materials. The intention is to extract the information around the atomic structure to create the neural network potential libraries used to train the neural network at the next stage. The scheme looks promising, but this idea is vividly influenced through the pseudo-potential approach of ab initio, first-principle Density Functional theory (DFT). Therefore there is a sprout acceleration in designing different descriptors in the DFT community that heavily invested in the density functional calculations for generations. \cite{PhysRevB.95.094203,ONG2013314,Hjorth_Larsen_2017,HIMANEN2020106949} Four generations of neural network potential descriptors developed and discretely existed in the literature in line with different types of Functional used in the DFT, from simple LDA to computationally expansive hybrid HSC Functional. Different atomistic and molecular descriptor fingerprinting configurations are available in the neural network potential, e.g., Coulomb matrix, \cite{PhysRevLett.108.058301} Ewald sum matrix, sine matrix, \cite{Faber_Crystal}  Many-body Tensor Representation (MBTR), \cite{huo2017unified} Atom-centered Symmetry Function (ACSF), weighted atom-centered symmetry functions (wACSF), and spin-ACSF, \cite{Behler_Atom-centered,doi:10.1063/1.5019667} Smooth Overlap of Atomic Positions (SOAP), \cite{PhysRevB.87.184115}, Gaussian Approximation Potentials (GAP), \cite{PhysRevLett.104.136403}, Moment Tensor Potentials (MTP), \cite{doi:10.1137/15M1054183}  Spectral Neighbor Analysis Potential (SNAP), \cite{THOMPSON2015316,PhysRevMaterials.1.043603,doi:10.1063/1.5017641}, electrostatic SNAP (eSNAP). \cite{deng2019electrostatic} Most of these descriptors follow Lennard-Jone's potential profile around the local atom and molecule structure. They have an exponential cutoff radius usually tuned to vary in the range of $ \sim $ 6-10 {\AA}. Continuous potential designs using a simple analytical expression and piecewise added in various cutoff radius to make plane waves of local orbitals moment. In this way, the DFT community early harvested the low-hanging fruits by designing the neural network potential to use deep learning techniques to predict material properties. Nevertheless, the excitement towards the modern Deep learning approach where network architecture and topological design embeds the underlying physical principle steadily increases. Various more powerful network topologies for the specific application will enrich the physical chemistry and materials science field. For example, a convolution neural network harnesses the power of convolution operation in image segmentation and analysis work.
Similarly, a recursive neural network has the inherent time-domain information imbibe in the network and deployed for a non-markovian process-based dataset predication where memory information of dynamics is crucial. Similarly, a Graph neural network (GNN) approach uses to predict the organic molecule crystal properties. The molecules made of different monomers ring geometries mapped as different nodes of the graph of the network. There are various ways where underline physics can be directly incorporated into the topology of a neural network to estimate various properties. The advantage of such neural network topologies is that they are descriptor-free and do not necessitate first-level feature engineering to fingerprint the local atomistic environment potential. Hence, they can realize the full potential of neural network prediction without an underline approximation bottleneck. In all the descriptor-based neural potential training, the initial domain of the learning dataset is fixed by the most commonly occurring element in organic molecules, e.g., Carbon, Hydrogen,  Oxygen, Nitrogen atoms. However, in nature, molecular materials are made by diversifying combinations of atoms and their orbital interaction. Hence at the first learning stage of descriptor fingerprint, restricting the learning dataset to a few known customized combinations of atomic configuration grossly handicapped the true potential of information inferencing capability from a deep neural network. While designing the novel neural network topologies, we should consider the following question carefully. How much information do we want to encode in the neural network channel of corresponding breadth and depth, i.e., hyperparameters optimization, and what constraints do we apply to the network topology. The correlation between the expansion of hyperparameters and accuracy of information estimate or quantification of uncertainty over the network. The node weights and biases in one batch on various hidden layers represent some meaningful information of significance or have some spatial or temporal hyperparameter's values distribution. How does the prediction influence if we cutoff portions of hyperparameters in a trained network, and how do loss functions change with a varying internal degree of freedoms? A neural network can be metaphorically analogous to Feynman's path trajectories, where there can be initially infinite paths similar to a deep network interconnect in various hidden layers. The propagator kernel can be equivalent to a convolution function operator between the network layers. The stochastic functional derivative obtains in the backpropagation cycle to get gradients of hyperparameters.
Furthermore, the Least Action functional in Feynman's to move a particle from one point to another along a trajectory is equivalent to minimizing the loss/cost function along with the network. From the statistical point of view, the probability weight of a particular trajectory that a particle can take in the ensemble of available trajectories does not have any significant interpretation or helpful information. Similarly, a particular weight, bias, and activation function configuration on a node in the hidden layer of a deep neural network for one batch of the dataset do not affect the overall prediction outcome of the network. In this regard, we will provide another analogy related to ideal gas dynamics following the maxwell distribution in the microscopic reversibility environment. In this configuration, if we sample the motion of the gas molecule in space-time. Following the trajectory for even infinite time does not yield us a piece of valuable information about the whole distribution of the gas particle in the chamber. The one advantage of physical dynamics informed network topology design is that we can determine and visualize the unknown constitutive relation between various fundamental properties of matter without solving the higher-order differential equation as they bottlenecked due to the curse of dimensionality. We can infer higher-order constitutive relation or momenta of the differential equation by parameterizing and mapping the elementary particle properties, e.g., charge, spin, and mass, on a deep neural network's $n$ input nodes. Such as inter-relation between charge, spin, heat current, conductivity, thermoelectric effect, and Peltier effect, without solving four equations of the first-order momenta of Boltzmann equation. Also, Less known inter-relation between hyperparameters can be visualized and estimated. Furthermore, a data-centric physical science approach can establish a novel correlation between hyperparameters.
We have the analytical expression for hyperparameter space's low order related properties based on nature's physical law. However, as deep neural networks did not suffer from the curse of dimensionality, we can extract more information from the training dataset and predict more inter-relation from learning. For a dissipative system connected with the external environment, fluctuation in generalized forces is related to generalized resistance in the path through the fluctuations-dissipation theorem. \cite{onsager_reciprocal_1931, onsager_reciprocal_1931-1,Kubo_1966} The Nyquist relation can be mapped and parameterized on the deep neural network to predict a dissipative system's electrical impedance and admittance. \cite{nyquist_thermal_1928,callen_irreversibility_1951}
A particular node of the deep neural network is analogous to the microcanonical state of a gas molecule, and the macroscopic grand canonical ensemble statistically distributes across the trained network in terms of learned biases, weights, and activation functions. Suppose a network is not learned enough by minimizing its cost and loss function after a specific epoch cycle from the training dataset in the learning stage. We can assume this is analogous to gas molecules in the chamber. There is not enough sampling time to distribute gas molecules in the volume space, and degrees of freedom did not interact sufficiently with the environment. We can safely assume that after sufficient time provided to gas molecules and afterward sampling, the statistical ensemble average of the microscopic states will exhibit in the macrostate similar to a neural network is trained to achieve the stationary or steady-state solution. This temporal evolution of gas molecules can speed up in the network by using batch mode of the parallel training process. It could be understood as we are performing the interaction experiments on multiple gas chambers with a very uniquely distinct initial microcanonical state of gas molecules in configuration space with the same degree of freedoms available in each gas chamber. Furthermore, the various training datasets for the learning rate optimization in the deep neural network are analogous to the different initial spatial distribution of gas chamber macrostates to start the gas molecules interaction experiment. As expected for a few skewed macrostates, it takes more time to reach the equilibrium optima, equivalent to more time a network needs to learn statistics and predict the unseen dataset.
Furthermore, we expect that if the input dataset fairly distributes in hyperparameter space similar to random spatially distributed gas molecules in configuration space, it will quickly achieve its steady-state ensemble statistics. We will conclude the phenomenological analogy about the deep neural network with the last circumspect animadversion. A deep neural network trained on a dataset which is statistically equivalent to all possible molecules distribution of gas chamber except the one unique distribution in which all the molecular particle concentration in one corner of the chamber, in the physical experimental domain, such an event outcome happen at the time scale of the universe. Similarly, without seeing this distinct data state in the learning stage, the neural network can never predict such a specific event, even going through an infinite long training cycle on all other infinite-state data distribution. The deep neural network will try to learn about this unique state forever because it has never been seen in a training state such as a unique cornered molecular gas distribution. Therefore, it is crucial to properly construct the query the loss and cost function and ask for prediction from the black-box deep neural network model while applying it in the physical science domain.

\bibliographystyle{IEEEtran}
\bibliography{Lagrangian_Density_Space-Time_DNN_Topology}
\end{document}